\documentclass[10pt,twocolumn,english,showpacs,showkeys,prb]{revtex4}
\usepackage[T1]{fontenc}
\usepackage[latin9]{inputenc}
\usepackage[active]{srcltx}
\usepackage{mathrsfs}
\usepackage{amsmath}
\usepackage{amssymb}
\usepackage{graphicx}

\makeatletter

\providecommand{\tabularnewline}{\\}

\@ifundefined{textcolor}{}
{%
 \definecolor{BLACK}{gray}{0}
 \definecolor{WHITE}{gray}{1}
 \definecolor{RED}{rgb}{1,0,0}
 \definecolor{GREEN}{rgb}{0,1,0}
 \definecolor{BLUE}{rgb}{0,0,1}
 \definecolor{CYAN}{cmyk}{1,0,0,0}
 \definecolor{MAGENTA}{cmyk}{0,1,0,0}
 \definecolor{YELLOW}{cmyk}{0,0,1,0}
 }

\@ifundefined{showcaptionsetup}{}{%
 \PassOptionsToPackage{caption=false}{subfig}}
\usepackage{subfig}
\makeatother

\usepackage{babel}
\begin{document}

\title{Geometrical frustration of an extended Hubbard diamond chain in the quasi-atomic
limit}

\author{Onofre Rojas$^{1}$, S. M. de Souza$^{1}$ and N. S. Ananikian$^{1,2}$ }

\affiliation{$^{1}$Departamento de Ciencias Exatas, Universidade Federal de Lavras,
CP 3037, 37200-000, Lavras- MG, Brazil}

\affiliation{$^{2}$A.I. Alikhanyan National Science Laboratory, 0036 Yerevan,
Armenia.}
\begin{abstract}
We study the geometrical frustration of extended Hubbard model on
diamond chain, where vertical lines correspond to the hopping and
repulsive Coulomb interaction terms between sites, while the rest of
them represent only the Coulomb repulsion one. The phase diagrams at
zero temperature show quite curious phases: five types of frustrated
states and four types of non-frustrated ones, ordered
antiferromagnetically. Although decoration transformation was
derived to spin coupling systems, this approach can be applied to
electron coupling ones. Thus the extended Hubbard model can be
mapped onto another effective extended Hubbard model in atomic limit
with additional three and four-body couplings. This effective model
is solved exactly through transfer matrix method. In addition, using
the exact solution of this model we discuss several thermodynamic
properties away from half filled band, such as chemical potential
behavior, electronic density, entropy, where we study the
geometrical frustration, consequently we investigate the specific
heat as well.
\end{abstract}



\maketitle

\section{\label{sec:intro}Introduction}

The Hubbard model is one of the simplest model for strongly
interacting electron systems. In general, rigorous analysis of the
model is a difficult task, only in particular case a number of
rigorous exact results have been obtained \cite{elliot}. On the
other hand, geometrical frustration in strongly correlated electron
systems have attracted a great deal of interest over the past
decades \cite{hagemann,lierop}. For instance, due to the competition
between the nearest and the next-nearest exchange coupling, the
inorganic spin-Peierls compound exhibits a transition from a gapless
phase to a gaped dimerized ground state \cite{eggert}. The quantum
phase transition point from the gapless phase to the gapped
dimerized phase of this model was first determined by Okamoto and
Nomura\cite{okamoto}.

 The interplay between geometrical
frustration and strong electron correlation results in a complicated
phase diagram, containing many interesting phases: spin-gaped
metallic phase, disorder magnetic insulation phase, Heisenberg
insulator \cite{laflorencie}. The Hubbard model on the triangular
lattice shows a transition from a paramagnetic metal to an
antiferromagnetic insulator as the Hubbard on-site repulsion
gradually grows \cite{daul}.

Using the numerical diagonalization of finite size system  Hida
predicted the appearance of
the magnetization plateau  in one of the pioneering theoretical papers \cite{Hida}.
In the
framework of the transfer-matrix and dynamical recursive approaches
the frustrated magnetization plateaus were obtained for
ferromagnetic-ferromagnetic-antiferro\-magnetic, kagome chains as well
as zigzag ladder with multi-spin exchanges
\cite{Ohanyan,Hovhan,Artuso}.

For small nanoclusters by the exact numerical diagonalization the average electron density, magnetization pla\-teaus via
chemical potential or magnetic field have been studied
in
Hubbard model \cite{Koch,Koch2}.

Recently geometrical frustration of Hubbard model was widely
studied, particularly the diamond chain structure such as considered
by Derzhko et al. \cite{Derzhko09,Derzhko10}, where they discuss the
frustration for a special class of lattice. Monte\-negro-Filho and
Coutinho-Filho \cite{montenegro06} also considered the doped
$AB_{2}$ Hubbard chain, both in the weak coupling and the
infinite-$U$ limit (atomic limit). They studied a quite interesting
phases as a function of hole doping away from half-filled band.
Mancini \cite{Mancini,mancini2008} has presented the exact solution
of  extended one-dimensional  Hubbard model in the atomic limit,
obtaining   the chemical potential plateaus of the particle density,
of the on-site potential at zero temperature and studied the
thermodynamic charge susceptibility, compressibility at finite
temperature, as well as  other physical quantities. Vidal et al.
\cite{Vidal98-00} also discussed two interacting particles evolving
in a diamond chain structure embedded in a magnetic field. As the
particles interact, it leads to the strong localization induced by
the magnetic field for the particular value of a flux. Analogous
model was also studied by Rossler and Mainemer \cite{rossler}. The
Hubbard model in other quasi-one-dimensional triangular structure
also was studied by Wang \cite{WZ Wang}. The latter indicated that
for small hopping term, the system exhibits short range
antiferromagnetic correlation, whereas, when the hopping terms
become greater than critical point, there is a transition from an
antiferromagnetic state to a frustrated one. Moreover, the
insulator-metal transition takes place at hopping interaction even
greater than another critical point. Further Gulacsi et al.
\cite{gulacsi,gulacsi PTP} also discussed the diamond Hubbard chain
in a magnetic field and  a wide range of properties such as
flat-band ferromagnetism and correlation-induced metallic,
half-metallic process. The spinless versions of the Hubbard model on
diamond chain was also recently investigated through exact
analytical solution\cite{spinless}, as well as Lopes and
Dias\cite{Lopes} performed a detailed investigation using the exact
diagonalization approach.

In the last decade several diamond chain structures were discussed.
Some of them were motivated by real materials such as
$\mathrm{Cu_{3}(CO_{3})_{2}(OH)_{2}}$ known as azurite, which is an
interesting quantum antiferromagnetic model,  described by
Heisenberg model on generalized diamond chain. Honecker et al.
\cite{honecker} studied the dynamic and thermodynamic properties for
this model. Besides, Pereira et al. \cite{Pereira08} investigate the
magnetization plateau of delocalized interstitial spins on diamond
chain, as well as  they detect magnetocaloric effect in kinetically
frustrated diamond chain \cite{pereira09}. Quite recently, Lisnii
\cite{lisnii} studied a distorted diamond Ising-Hubbard chain and
that model exhibits geometrical frustration also. A further
investigation regarding the exact evidence for the spontaneous
antiferromagnetic long-range order in the two-dimensional hybrid
model of localized Ising spins and itinerant electrons was discussed
by Stre\v{c}ka et al. \cite{strecka prb,strecka conf}. Moreover, the
thermodynamics of the Ising-Heisenberg model on diamond-like chain
was widely discussed in references
\cite{canova06,vadim,valverde,lisnii-11}, and also the thermal
entanglement was considered by Ananikian et al. \cite{ananikian}.

On the other hand, the analytical exact solution is rather amazing,
since the exact result is always useful to manipulate against the
numerical results. Therefore, the exact solutions of the models are
of great importance, so our main goal is to obtain an exact solution
for the extended Hubbard model on diamond chain in quasi-atomic
limit.  The research of the extended Hubbard diamond chain model
without the hopping of electrons between the nodal sites is based on
the three main reasons. First of all 1/3 magnetization plateau, the
double peaks both in the magnetic susceptibility and specific heat
was observed in the experimental
measurements\cite{rule,kikuchi}according to the experiments of the
natural mineral azurite. Theoretical calculations of the
one-dimensional Hubbard model, as well as the experimental result of
the exchange dimer (interstitial sites) parameter and their
descriptions of the various theoretical models are presented. It
should be noted that the dimers (interstitial sites) exchange much
more strongly than those nodal sites. There were proposed various
types of theoretical Heisenberg approximate methods: the
renormalization of the density matrix renormalization group of the
transfer matrix, the density functional theory, the high temperature
expansion, Lanczos diagonalization on a diamond chain are to explain
the experimental measurements (magnetization plateau and the double
top) in natural mineral azurite\cite{honecker-lauchli}. All of these
theoretical studies are approximate. There is another possibility.
Since dimer interaction is much higher than the rest, it can be
represented as an exactly solvable Ising-Heisenberg model. In
addition, experimental data on the magnetization plateau coincide
with the approximation of Ising-Heisenberg model
\cite{canova06,Pereira08,ananikian,chakh} and the extended Hubbard
model without the hopping of electrons between the nodal
sites\cite{spinless} on a diamond chain.  This is the first reason.

The second one is  the quantum block-block entanglement, carried out by  exact diagonalization technique in one-dimensional extended Hubbard model, calculated for finite size (L=10)\cite{sh-deng}. When the absolute value of nearest-neighbor Coulomb interactions (our case)   becomes smaller, the effect of the hoping term and the on-site interaction cannot be neglected. Finally, we would like to point out that although the results, obtained in this paper are for the  2-site (dimer) system, their qualitative features are the same as for  big size system ones.

And, the third reason is the experimental observation of the double peaks both in the magnetic susceptibility and specific heat \cite{Pereira08,honecker-lauchli,Bo-Gu} may be described exactly by the extended Hubbard diamond chain model without the hopping of electrons or holes between the nodal sites. This phenomenon
is particularly important in the quantum case.

This paper is organized as follows: in Sec.~\ref{sec:Extended} we present the
extended Hubbard model on diamond chain.
 We have studied the phase diagram at zero temperature  in Sec.~\ref{sec:phase}.
Further, in Sec.~\ref{sec:Exact}, we present the exact solution of the model. In
Sec.~\ref{sec:Therm}, we have discussed the thermodynamic properties of the model,
such as electronic density, internal energy, compressibility,
entropy and specific heat away from the half-filled band. Finally,
Sec.~\ref{sec:conclusions} contains the concluding remarks.

\section{\label{sec:Extended}The Extended Hubbard model}

The extended Hubbard model on diamond chain will be
considered in this paper, as represented
schematically in FIG.~\ref{fig:extd-hubbrd-diamd}. In the present model we consider the
hopping term $t$ between sites $a$ and $b$, besides, there are
onsite Coulomb repulsion interaction term denoted by $U$ and the nearest
neighbor repulsion interaction term, denoted by $V$. We assume also
that this model has an arbitrary particle density, so, we will
include a chemical potential term noted by $\mu$, therefore the
Hamiltonian of this model can be expressed by:
\begin{figure}
\includegraphics[scale=0.50]{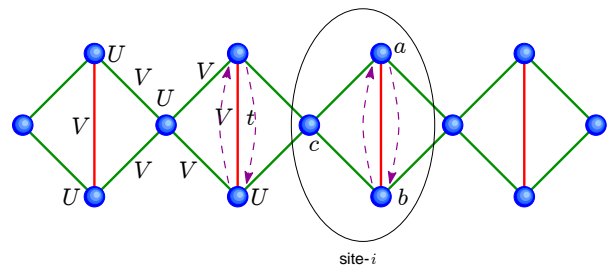}\caption{\label{fig:extd-hubbrd-diamd}(Color online) Schematic representation of extended
Hubbard model on diamond chain.}
\end{figure}
\begin{equation}
\boldsymbol{H}=\sum_{i=1}^{N}\boldsymbol{H}_{i,i+1},
\end{equation}
with $N$ being the number of cell (sites $a$, $b$ and $c$), whereas $\boldsymbol{H}_{i,i+1} $ is given by

\begin{equation}
\begin{array}{l}
\boldsymbol{H}_{i,i+1}= -t\sum\limits_{\sigma=\downarrow,\uparrow}\left(\boldsymbol{a}_{i,\sigma}^{\dagger}\boldsymbol{b}_{i,\sigma}+\boldsymbol{a}_{i,\sigma}\boldsymbol{b}_{i,\sigma}^{\dagger}\right)-
\\
-\ \mu\left(\boldsymbol{n}_{i}^{a}+\boldsymbol{n}_{i}^{b}+\tfrac{1}{2}(\boldsymbol{n}_{i}^{c}+\boldsymbol{n}_{i+1}^{c})\right)+\\
+\ U\left(\boldsymbol{n}_{i,\uparrow}^{a}\boldsymbol{n}_{i,\downarrow}^{a}+\boldsymbol{n}_{i,\uparrow}^{b}\boldsymbol{n}_{i,\downarrow}^{b}+\tfrac{1}{2}(\boldsymbol{n}_{i,\uparrow}^{c}\boldsymbol{n}_{i,\downarrow}^{c}+\boldsymbol{n}_{i+1,\uparrow}^{c}\boldsymbol{n}_{i+1,\downarrow}^{c})\right)+\\
+\ V\left(\boldsymbol{n}_{i}^{a}\boldsymbol{n}_{i}^{b}+\tfrac{1}{2}(\boldsymbol{n}_{i}^{a}+\boldsymbol{n}_{i}^{b})(\boldsymbol{n}_{i}^{c}+\boldsymbol{n}_{i+1}^{c})\right),\label{eq:Hamt-dmnd}
 \end{array}
\end{equation}
with $\boldsymbol{a}_{i,\sigma}$, and $\boldsymbol{b}_{i,\sigma}$
($\boldsymbol{a}_{i,\sigma}^{\dagger}$ and $\boldsymbol{b}_{i,\sigma}^{\dagger}$)
being the Fermi annihilation (creation) operator for the Hubbard model,
while by $\sigma$ we represent the electron spin, and by $\boldsymbol{n}_{i,\sigma}^{\alpha}$ we mean the number operator, with $\alpha=\{a,b,c\}$, using this number operator we define also the following operators $\boldsymbol{n}_{i}^{\alpha}=\boldsymbol{n}_{i,\uparrow}^{\alpha}+\boldsymbol{n}_{i,\downarrow}^{\alpha}$.

In order to contract the Hamiltonian (\ref{eq:Hamt-dmnd}), we can
define  the following operators
\begin{equation}
\begin{array}{rcl}
\boldsymbol{p}_{i,i+1} & =&\tfrac{1}{2}(\boldsymbol{n}_{i}^{c}+\boldsymbol{n}_{i+1}^{c}),\\
\boldsymbol{q}_{i,i+1} & =&\tfrac{1}{2}(\boldsymbol{n}_{i,\uparrow}^{c}\boldsymbol{n}_{i,\downarrow}^{c}+\boldsymbol{n}_{i+1,\uparrow}^{c}\boldsymbol{n}_{i+1,\downarrow}^{c}),
\end{array}
\end{equation}

using  these operators, we can rewrite the Hamiltonian (\ref{eq:Hamt-dmnd}),
as follows:
\begin{equation}
\begin{array}{ccl}
\boldsymbol{H}_{i,i+1}&= & -t\sum\limits_{\sigma=\downarrow,\uparrow}\left(\boldsymbol{a}_{i,\sigma}^{\dagger}\boldsymbol{b}_{i,\sigma}+\boldsymbol{a}_{i,\sigma}\boldsymbol{b}_{i,\sigma}^{\dagger}\right)-\\
&-&(\mu-V\boldsymbol{p}_{i,i+1})\left(\boldsymbol{n}_{i}^{a}+\boldsymbol{n}_{i}^{b}\right)+ \\
 & +&U\left(\boldsymbol{n}_{i,\uparrow}^{a}\boldsymbol{n}_{i,\downarrow}^{a}+\boldsymbol{n}_{i,\uparrow}^{b}\boldsymbol{n}_{i,\downarrow}^{b}\right)+\\
 &+&V\boldsymbol{n}_{i}^{a}\boldsymbol{n}_{i}^{b}-\mu\boldsymbol{p}_{i,i+1}+U\boldsymbol{q}_{i,i+1}.\label{eq:Ham-red}
\end{array}
\end{equation}

The symmetry particle-hole can be analyzed in a similar way, as discussed
in reference \cite{spinless}. The phase diagram in the half-filled
band, can be obtained using the following restriction for the chemical
potential $\mu=U/2+2V$. This relation could be obtained by using the
symmetry particle-hole in a similar way, as discussed in reference\cite{spinless}.

\section{\label{sec:phase}The phase diagram}

In order to study the phase diagram at zero temperature, first we
need to diagonalize the Hamiltonian (\ref{eq:Ham-red}) at sites $a$
and $b$.  Where by $m=\{0,..,4\}$ we will denote the total number of mobile electrons per unit cell at sites $a$ and $b$, while by $n_{c}$ we mean the number of electrons per unit cell at site-c.
The Hamiltonian at sites $a$ and $b$ can be written as $16\times16$ matrix, but this matrix
can be constructed as a block matrix, where in the largest block
matrix we have $4\times4$, therefore the eigenvalues and eigenvectors of this matrix
are expressed below as:
\begin{description}
\item [{(a)}] State with $m=0$ particle
\begin{equation}
\lambda_{0}=\boldsymbol{D}_{0},\quad|S_{0}\rangle=|0,0\rangle,
\end{equation}
where $\boldsymbol{D}_{m}=-(\mu-V\boldsymbol{p}_{i,i+1})m-\mu\boldsymbol{p}_{i,i+1}+U\boldsymbol{q}_{i,i+1}$,
 While the state $\quad|S_{0}\rangle=|0,0\rangle$ means there are no particle at site $a$ and $b,$ respectively.
\item [{(b)}] State with $m=1$ particle
\begin{equation}
\lambda_{\sigma}^{(\pm)}=\boldsymbol{D}_{1}\pm t,\quad|S_{\sigma}^{(\pm)}\rangle=\tfrac{1}{\sqrt{2}}\left(|0,\sigma\rangle\mp|\sigma,0\rangle\right),
\end{equation}
with $|S_{\sigma}^{(\pm)}\rangle$ we mean there is one particle at site $a$ or $b$ with either spin up or down.
\item [{(c)}] State with $m=2$ particles
\begin{align}
\lambda_{2\sigma} & =\boldsymbol{D}_{2}+V,\quad|S_{2\sigma}\rangle=|\sigma,\sigma\rangle,\\
\lambda_{\downarrow\negmedspace\uparrow}^{(1)} & =\boldsymbol{D}_{2}+U,\quad|S_{\downarrow\negmedspace\uparrow}^{(1)}\rangle=\tfrac{1}{\sqrt{2}}\left(|\negmedspace\downarrow\negmedspace\uparrow,0\rangle-|0,\negmedspace\downarrow\negmedspace\uparrow\rangle\right),\\
\lambda_{\downarrow\negmedspace\uparrow}^{(\pm)} & =\boldsymbol{D}_{2}+V+\frac{\theta_{\pm}t}{2},\\
|S_{\downarrow\negmedspace\uparrow}^{(\pm)}\rangle & =\tfrac{1}{\zeta_{\mp}}\left(|\negmedspace\downarrow\negmedspace\uparrow,0\rangle+|0,\negmedspace\downarrow\negmedspace\uparrow\rangle+\theta_{\mp}\left(|\uparrow,\downarrow\rangle+|\downarrow,\uparrow\rangle\right)\right)\\
\lambda_{\downarrow\negmedspace\uparrow}^{(2)} & =\boldsymbol{D}_{2}+V,\quad|S_{\downarrow\negmedspace\uparrow}^{(2)}\rangle=\tfrac{1}{\sqrt{2}}\left(|\uparrow,\downarrow\rangle-|\downarrow,\uparrow\rangle\right),
\end{align}
where
\begin{align}
\zeta_{\pm}= & \sqrt{2+\tfrac{\theta_{\pm}^{2}}{8}},\\
\theta_{\pm}= & \frac{U-V\pm\sqrt{(U-V)^{2}+16t^{2}}}{t},\label{thetas}
\end{align}
the states  $|S_{2\sigma}\rangle$ and $|S_{\downarrow\negmedspace\uparrow}^{(\pm)}\rangle$ are defined in a similar way as defined in the previous case, two particles both with spin up or down and two particles with opposite spins.
\item [{(d)}] State with $m=3$ particles
\begin{equation}
\lambda_{\downarrow\negmedspace\uparrow\sigma}^{(\pm)}=\boldsymbol{D}_{3}+U+2V\pm t,\quad|S_{\downarrow\negmedspace\uparrow\sigma}^{(\pm)}\rangle=\tfrac{1}{\sqrt{2}}\left(|\negmedspace\downarrow\negmedspace\uparrow,\sigma\rangle\mp|\sigma,\negmedspace\downarrow\negmedspace\uparrow\rangle\right),
\end{equation}
the states $|S_{\downarrow\negmedspace\uparrow\sigma}^{(\pm)}\rangle$ correspond to two particles with  opposite spin and one spin
\item [{(e)}] State with $m=4$ particles
\begin{equation}
\lambda_{\downarrow\negmedspace\uparrow\downarrow\negmedspace\uparrow}=\boldsymbol{D}_{4}+2U+4V,\quad|S_{\downarrow\negmedspace\uparrow\downarrow\negmedspace\uparrow}\rangle=|\negmedspace\downarrow\negmedspace\uparrow,\downarrow\negmedspace\uparrow\rangle.
\end{equation}
the state
$|S_{\downarrow\negmedspace\uparrow\downarrow\negmedspace\uparrow}\rangle$
means there are two spins at each site with opposite spins.
\end{description}
It is worth to note that, the Hamiltonian (\ref{eq:Hamt-dmnd}),
has 256 eigenvalues per diamond plaquette.

From here after, we will consider only repulsive onsite Coulomb interaction ($U>0$ ) as well as repulsive Coulomb interaction between nearest neighbor  ($V>0$). Besides $V<U$, in order to make more realistic our discussion. 

\begin{figure}
\includegraphics[angle=-90,scale=0.35]{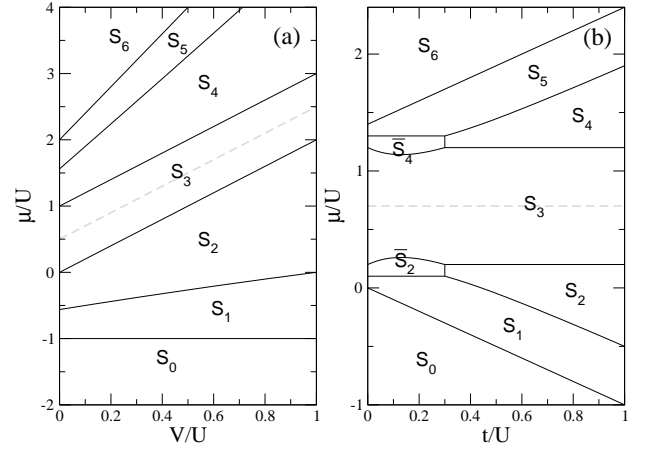}\caption{\label{fig-diamg}Phase diagrams at zero temperature: (a) displays
the phase diagram of $V/U$ versus $\mu/U$ for fixed value of $t/U=1$.
Whereas in (b) displays the phase diagram of $t/U$ versus $\mu/U$ for fixed
value of $V/U=0.25$.}
\end{figure}

Afterwards we will discuss the phase diagram at zero temperature,
the present model exhibits 9 states which are tabulated in table \ref{grnd-st},
for all possible numbers of particles available in the chain.  For the
empty particle and fully filled particles, we have one state $|\mathsf{S_{0}\rangle=|Vac\rangle}$
and $|\mathsf{S_{6}\rangle=|Full\rangle,}$ respectively. While for
one particle per unit cell and five particles per unit cell, we also
have one corresponding frustrated state $|\mathsf{S_{1}\rangle=|FRU_{1}\rangle}$
and $|\mathsf{S_{5}\rangle=|FRU_{5}\rangle,}$ respectively. However,
for 2 particles per unit cell, we have two states, one configuration
is an antiferromagnetic state ($\mathsf{|S_{2}\rangle=|AFM_{2}\rangle}$),
whereas other configuration is a frustrated state ($\mathsf{|\bar{S}_{2}\rangle=|FRU_{2}\rangle}$).
Due to particle-hole symmetry the analyze for the case of four particles
(or two holes) becomes analogous for the case of two particles, hence,
we have one antiferromagnetically ordered state ($\mathsf{|S_{4}\rangle=|AFM_{4}\rangle}$)
and another frustrated state ($\mathsf{|\bar{S}_{4}\rangle=|FRU_{4}\rangle}$).
Furthermore, for the special case of half-filled particles (with 3
particles or holes per unit cell) we also could have two possible
states, one frustrated state $\mathsf{|S_{3}\rangle=|FRU_{3}\rangle}$
and another antiferromagnetically ordered state ($|\bar{\mathsf{S}}_{3}\rangle=|\mathsf{AFM_{3}\rangle}$),
however the case $|\mathsf{AFM_{3}}\rangle$ occurs only when $V>U,$
this case we  ignore, since, it is too artificial one.
\begin{table}
\begin{tabular}{|c|c|c|}
\hline
$n_{t}$ & Ground state energy & Ground state\tabularnewline
\hline
\hline
0 & $0$ & $\mathsf{|S_{0}\rangle=}|\mathsf{Vac}\rangle=\prod_{i}|S_{0}\rangle_{i}\otimes|0\rangle_{i}$\tabularnewline
\hline
1 & $-\mu-t$ & $|\mathsf{S_{1}\rangle=}|\mathsf{FRU}_{1}\rangle=\prod_{i}|S_{\sigma}^{(-)}\rangle_{i}\otimes|0\rangle_{i}$\tabularnewline
\hline
2 & $-2\mu+V+\frac{\theta_{-}t}{2}$ & $|\mathsf{S_{2}\rangle=}|\mathsf{AFM}_{2}\rangle=\prod_{i}|S_{\downarrow\!\uparrow}^{(-)}\rangle_{i}\otimes|0\rangle_{i}$\tabularnewline
\hline
2 & $-2\mu-t+V$ & $|\mathsf{\bar{S}_{2}\rangle=}|\mathsf{FRU}_{2}\rangle=\prod_{i}|S_{\sigma}^{(-)}\rangle_{i}\otimes|1\rangle_{i}$\tabularnewline
\hline
3 & $-3\mu+3V+\frac{\theta_{-}t}{2}$ & $|\mathsf{S_{3}\rangle=}|\mathsf{FRU}_{3}\rangle=\prod_{i}|S_{\downarrow\!\uparrow}^{(-)}\rangle_{i}\otimes|1\rangle_{i}$\tabularnewline
\hline
4 & $-4\mu+5V+U+\frac{\theta_{-}t}{2}$ & $|\mathsf{S_{4}\rangle=}|\mathsf{AFM}_{4}\rangle=\prod_{i}|S_{\downarrow\!\uparrow}^{(-)}\rangle_{i}\otimes|2\rangle_{i}$\tabularnewline
\hline
4 & $-4\mu-t+5V+U$ & $|\mathsf{\bar{S}_{4}\rangle=}|\mathsf{FRU}_{4}\rangle=\prod_{i}|S_{\downarrow\!\uparrow\sigma}^{(-)}\rangle_{i}\otimes|1\rangle_{i}$\tabularnewline
\hline
5 & $-5\mu+8V+2U-t$ & $|\mathsf{S_{5}\rangle=}|\mathsf{FRU}_{5}\rangle=\prod_{i}|S_{\downarrow\!\uparrow\sigma}^{(-)}\rangle_{i}\otimes|2\rangle_{i}$\tabularnewline
\hline
6 & $-6\mu+12V+3U$ & $|\mathsf{S_{6}\rangle=}|\mathsf{Full}\rangle=\prod_{i}|S_{\downarrow\!\uparrow\downarrow\!\uparrow}\rangle_{i}\otimes|2\rangle_{i}$\tabularnewline
\hline
\end{tabular}
\caption{\label{grnd-st}The first column $n_t$ represents the number of particles
per unit cell, in the second and third column is tabulated the ground
state energy per unit cell and its corresponding ground state respectively.}
\end{table}
 In FIG.~\ref{fig-diamg} we illustrate the phase transition at
zero temperature of these states. In FIG.~\ref{fig-diamg}(a) we
illustrate the phase diagram of $V/U$ versus $\mu/U$, for fixed
value of $t/U=1$, where we display seven states $\mathsf{S}_{i}$
($i=0..6$). While in FIG.~\ref{fig-diamg}(b) we display the phase
diagram of $t/U$ versus $\mu/U$, for fixed value of $V/U=0.25$. This
phase diagram illustrates 9 states, the seven states are already
shown in FIG.~\ref{fig-diamg}(a) and two additional ones
$|\mathsf{FRU}_{2}\rangle$ and $|\mathsf{FRU}_{4}\rangle$. The fully
transition boundary functions between states are tabulated in table
\ref{bound-sta}. The first and third columns correspond to the
boundary states, while the second and fourth columns correspond to
the boundary functions.

In FIG.~\ref{fig-diamg}, by dashed line we represent the half-filled
band curve of the extended Hubbard model on diamond chain.

\begin{table}
\begin{tabular}{|c|c|c|c|}
\hline
State &  $V'\times\mu/U$ (fig. 2a) & State &  $t'\times\mu/U$ (fig. 2b)\tabularnewline
\hline
\hline
$\mathsf{S_{0},S_{1}}$ & -1 & $\mathsf{S_{0},S_{1}}$ & $-t'$\tabularnewline
\hline
$\mathsf{S_{1},S_{2}}$ & $\frac{3+V'-\sqrt{(V'-1)^{2}+16}}{2}$ & $\mathsf{S_{1},S_{2}}$ & $\frac{11-10\sqrt{0.81+16t'^{2}}+20t'}{20}$\tabularnewline
\hline
$\mathsf{S_{2},S_{3}}$ & $2V'$ & $\mathsf{S_{2},S_{3}}$ & $\tfrac{1}{5}$\tabularnewline
\hline
$\mathsf{S_{3},S_{4}}$ & $2V'+1$ & $\mathsf{S_{3},S_{4}}$ & $\frac{6}{5}$\tabularnewline
\hline
$\mathsf{S_{4},S_{5}}$ & $\frac{-1+7V'-\sqrt{(V'-1)^{2}+16}}{2}$ & $\mathsf{S_{4},S_{5}}$ & $\frac{17-20t'+5\sqrt{0.81+16t'^{2}}}{20}$\tabularnewline
\hline
$\mathsf{S_{5},S_{6}}$ & $4V'+2$ & $\mathsf{S_{5},S_{6}}$ & $t'+\tfrac{7}{5}$\tabularnewline
\hline
 &  & $\mathsf{S_{5},\bar{S}_{4}}$ & $\tfrac{13}{10}$\tabularnewline
\hline
 &  & $\mathsf{\bar{S}_{4},S_{3}}$ & $\frac{-4t'+3+2\sqrt{0.81+16t'^{2}}}{4}$\tabularnewline
\hline
 &  & $\mathsf{S_{3},\bar{S}_{2}}$ & $\frac{20t'+13-10\sqrt{0.81+16t'^{2}}}{20}$\tabularnewline
\hline
 &  & $\mathsf{\bar{S}_{2},S_{1}}$ & $\frac{1}{10}$\tabularnewline
\hline
\end{tabular}

\caption{\label{bound-sta}The first and third columns represent the boundary
between two states tabulated in table \ref{grnd-st}, while the second
and fourth columns mean the function of boundaries.  For simplicity we use the following notation $V'=V/U$ and $t'=t/U$}
\end{table}

Other variants of the phase diagram at zero temperature could also
be discussed, although the main properties already had been
discussed in FIG.~\ref{fig-diamg}.

\section{\label{sec:Exact}Exact solution}

The method to be used will be the decoration transformation early
proposed by Syozi \cite{Syozi} and Fisher \cite{Fisher}, more later
this approach was the subject of study by Rojas~et~al.~\cite{phys-A-09},
for the case of multi-spins. As well as by Stre\v{c}ka \cite{strecka pla}
for the  hybrid system (e.g. Ising-Heisenberg). Another interesting
variant of this approach is also  discussed by us, where we propose
a direct transformation instead of several step by step one \cite{JPA-11},
an illustrative successful application of the last approach was performed
in reference \cite{m rojas}. The decoration transformation approach
is widely used to solve spin models, however the decoration transformation
approach can also  be applied to electron coupling system, such have been
used by the authors \cite{spinless} for the case of spinless fermion
on diamond structure.

Therefore, in order to use the decoration transformation approach
we can write the Boltzmann factor for extended Hubbard model on diamond
chain as follows:
\begin{equation}
\begin{array}{l}
w_{\boldsymbol{n}_{i}^{c},\boldsymbol{n}_{i+1}^{c}}= \displaystyle \mathrm{e}^{-\beta\boldsymbol{D}_{0}}+4\left(\mathrm{e}^{-\beta\boldsymbol{D}_{1}}+\mathrm{e}^{-\beta(\boldsymbol{D}_{3}+U+2V)}\right)\cosh(\beta t)\\
\qquad +\displaystyle\mathrm{e}^{-\beta\boldsymbol{D}_{2}}\left(\mathrm{e}^{-\beta U}+3\mathrm{e}^{-\beta V}\right) +\mathrm{e}^{-\beta(\boldsymbol{D}_{4}+2U+4V)}+\\
\qquad +\displaystyle\mathrm{e}^{-\beta(\boldsymbol{D}_{2}+V)}\left(\mathrm{e}^{-\beta\theta_{+}t/2}+\mathrm{e}^{-\beta\theta_{-}t/2}\right),\label{eq:orig-W}
\end{array}
\end{equation} where  the $\theta_{\pm}$ already were defined in equation (\ref{thetas}).

The extended Hubbard model on diamond chain can be mapped onto an
effective extended Hubbard-like model in atomic limit, where it involves
additional three-body and four-body interaction. Therefore, the effective
extended Hubbard model becomes like:
\begin{equation}
\begin{array}{ccl}
\tilde{\boldsymbol{H}}_{i,i+1}&= & -\tilde{\mu}\boldsymbol{n}_{i}^{c}+\tilde{U}\boldsymbol{n}_{i,\uparrow}^{c}\boldsymbol{n}_{i,\downarrow}^{c} \\
 & +&\tilde{V}\left(\boldsymbol{n}_{i,\uparrow}^{c}+\boldsymbol{n}_{i,\downarrow}^{c}\right)\left(\boldsymbol{n}_{i+1,\uparrow}^{c}+\boldsymbol{n}_{i+1,\downarrow}^{c}\right) \\
 & +&\tilde{W}_{3}\boldsymbol{n}_{i,\uparrow}^{c}\boldsymbol{n}_{i,\downarrow}^{c}\left(\boldsymbol{n}_{i+1,\uparrow}^{c}+\boldsymbol{n}_{i+1,\downarrow}^{c}\right)\\
 & +&\tilde{W}_{4}\boldsymbol{n}_{i,\uparrow}^{c}\boldsymbol{n}_{i,\downarrow}^{c}\boldsymbol{n}_{i+1,\uparrow}^{c}\boldsymbol{n}_{i+1,\downarrow}^{c},
\end{array}
\label{eq:eff-Hamt}
\end{equation}
where $\tilde{\mu}$ could be interpreted as effective chemical potential,
in a similar way,  $\tilde{U}$ labels the onsite Coulomb repulsion coupling, meanwhile $\tilde{V}$ corresponds to the nearest neighbor repulsion coupling, the next terms,  $\tilde{W}_{3}$ will be interpreted
as three body interaction term, whereas $W_{4}$ as four body coupling. All the above parameters could be obtained as a function of Hamiltonian (\ref{eq:Hamt-dmnd}) parameters, when decoration transformation is performed

The Boltzmann factor for effective Hubbard model in atomic limit becomes
like:

\begin{equation}
\tilde{w}_{\boldsymbol{n}_{i}^{c},\boldsymbol{n}_{i+1}^{c}}=A\,\mathrm{e}^{-\beta\tilde{\boldsymbol{H}}_{i,i+1}},\label{eq:eff-W}
\end{equation}
the factor $A$ is an additional variable to be determined in terms
of the original Hamiltonian (\ref{eq:Hamt-dmnd}). By the use of
decoration transformation\cite{Syozi,Fisher,phys-A-09} we will
impose the following condition

\begin{equation}
w_{\boldsymbol{n}_{i}^{c},\boldsymbol{n}_{i+1}^{c}}=\tilde{w}_{\boldsymbol{n}_{i}^{c},\boldsymbol{n}_{i+1}^{c}},\label{eq:ff-cond}
\end{equation}
using Eq.~(\ref{eq:ff-cond}) we can find all parameters of the
effective Hamiltonian (\ref{eq:eff-Hamt}) and the factor $A$.

To solve the effective Hubbard model with up to four-body coupling,
we can use the transfer matrix method\cite{baxter-book}, similarly
to that used in reference \cite{spinless,beni-pincus}. Therefore
we symmetrize the Hamiltonian by exchanging $i\rightarrow i+1$ and
$i+1\rightarrow i$, thus, the transfer matrix becomes symmetric,
for our case, this transfer matrix could be expressed by:

\begin{equation}
{\bf T}=\left[\begin{array}{cccc}
w_{0,0} & w_{0,1} & w_{0,1} & w_{0,2}\\
w_{0,1} & w_{1,1} & w_{1,1} & w_{1,2}\\
w_{0,1} & w_{1,1} & w_{1,1} & w_{1,2}\\
w_{0,2} & w_{1,2} & w_{1,2} & w_{2,2}
\end{array}\right],\label{eq:transf-Mtx}
\end{equation}
where the Boltzmann factor is given by Eq.~(\ref{eq:orig-W}), and
using a convenient notation, such as
\begin{equation}
\begin{array}{ccl}
w_{0,0}(x)&= &\displaystyle 1+2x\left(1+\frac{x^{2}}{z^{4}y^{2}}\right)\left(\frac{1}{\gamma^{2}}+\gamma^{2}\right)+ \\
 & +&\displaystyle x^{2}\left(\frac{3}{z^{2}}+\frac{1}{y^{2}}+\frac{1}{yz\varsigma}+\frac{\varsigma}{yz}\right)+\frac{x^{4}}{y^{4}z^{8}},
\end{array}\label{eq:w00}
\end{equation}
 with $ x=\mathrm{e}^{2\beta\mu}$, $y=\mathrm{e}^{\frac{1}{2}\beta U}$,
$z=\mathrm{e}^{\frac{1}{2}\beta V}$,
$\gamma=\mathrm{e}^{\frac{1}{2}\beta t}$ and $
\varsigma=\mathrm{e}^{\frac{1}{2}\beta\sqrt{(U-V)^{2}+16t^{2}}}$.
Whereas the remaining Boltzmann factors could be expressed easily
through the function $w_{0,0}(x)$ defined in Eq.~(\ref{eq:w00}) as
follows:
\begin{equation}
w_{n_{1},n_{2}}(x)= \frac{x^{(n_{1}+n_{2})/2}}{y^{[n_{1}/2]+[n_{2}/2]}}w_{0,0}\left(\frac{x}{z^{n_{1}+n_{2}}}\right),
\end{equation}
 by $[\ldots]$ inside
braces we mean the least integer of any real number. Thereafter, all elements of matrix (\ref{eq:transf-Mtx}) are
well expressed just in terms of $w_{0,0}(x)$.

 The Boltzmann factors for the effective Hubbard model
(\ref{eq:eff-W}) with three and four-body terms are given by:

\begin{equation}
\begin{array}{cl}
\tilde{w}_{0,0}= & A,\\
\tilde{w}_{0,1}= & Ar;\quad r=\mathrm{e}^{\beta\tilde{\mu}/2},\\
\tilde{w}_{0,2}= & Ars;\quad s=\mathrm{e}^{-\beta\tilde{U}/2},\\
\tilde{w}_{1,1}= & Ar^{2}t;\quad t=\mathrm{e}^{-\beta\tilde{V}},\\
\tilde{w}_{1,2}= & Ar^{3}st^{2}u;\quad u=\mathrm{e}^{-\beta\tilde{W}_{3}/2},\\
\tilde{w}_{2,2}= & Ar^{4}s^{2}t^{4}u^{4}v.
\end{array}
\end{equation}

The determinant of transfer matrix becomes a quartic equation of the
form $\mathrm{det}({\bf T}-\Lambda)=\Lambda^{4}+a_{3}\Lambda^{3}+a_{2}\Lambda^{2}+a_{1}\Lambda$,
where the coefficients become:
\begin{equation}
\begin{array}{rcl}
a_{1}&=&-2w_{0,0}w_{1,1}w_{2,2}+2w_{0,2}^{2}w_{1,1}+2w_{0,1}^{2}w_{2,2}+\\
 &&+2w_{0,0}w_{1,2}^{2}-4w_{0,2}w_{0,1}w_{1,2},\\
a_{2}&=&-2w_{0,1}^{2}+2w_{0,0}w_{1,1}+w_{0,0}w_{2,2}+\\
 &&+2w_{1,1}w_{2,2}-w_{0,2}^{2}-2w_{1,2}^{2},\\
a_{3}&=&-w_{0,0}-w_{2,2}-2w_{1,1},
\end{array}
\end{equation}
or alternatively the coefficients also can be expressed, using the
effective Hamiltonian parameters, thus, the coefficients of the quartic
equation are given by:

\begin{equation}
\begin{array}{cl}a_{1}= &  A\left(-2r^{6}t^{5}s^{2}u^{4}v+2r^{6}s^{2}t+2r^{6}s^{2}t^{4}u^{4}v+\right.\\
 & \left.+2r^{6}s^{2}t^{4}u^{2}-4r^{6}s^{2}t^{2}u\right),\\
a_{2}= & A\left(-2r^{2}+2r^{2}t+r^{4}s^{2}t^{4}u^{4}v+2r^{6}t^{5}s^{2}u^{4}v-\right.\\
 & \left.-r^{4}s^{2}-2r^{6}s^{2}t^{4}u^{2}\right),\\
a_{3}= & -A\left(1+r^{4}s^{2}t^{4}u^{4}v+2r^{2}t\right).
\end{array}
\end{equation}

Therefore the roots of algebraic quartic equation may be reduced to
a cubic one,  solutions of which are given in the following way:
\begin{equation}
\Lambda_{j}=2\sqrt{Q}\cos\left(\tfrac{\phi+2\pi j}{3}\right)-\frac{1}{3}a_{3}\quad j=0,1,2,\label{eq:sol-cub}
\end{equation}
with
\begin{equation}
\begin{array}{ccl}\phi&=&  \arccos\left(\tfrac{R}{\sqrt{Q^{3}}}\right),\\
Q&= & \displaystyle\frac{a_{3}^{2}-3a_{2}}{9},\\
R&= & \displaystyle\frac{9a_{2}a_{3}-27a_{1}-2a_{3}^{3}}{54}.
\end{array}
\end{equation}
Furthermore, we also have additional trivial solution $\Lambda_{3}=0$,
of the algebraic quartic equation.

Hence, the largest eigenvalue of the transfer matrix will be
$\Lambda_{0}$, which is  expressed by  Eq.~(\ref{eq:sol-cub}). Once
known the largest eigenvalue of transfer matrix, we are able to
obtain
 the partition function straightforwardly, and thereafter the free
energy is given by $f=-\frac{1}{\beta}\ln(\Lambda_{0})$. Using the
free energy per unit cell, we may obtain the thermodynamic
properties and how the model behaves when the number particles are
changing away from the half filled band.

\section{\label{sec:Therm}Thermodynamic properties}

In order to study the thermodynamic properties we will use the exact
free energy $\displaystyle f=-\frac{1}{\beta}\ln(\Lambda_{0})$ as a starting point.
Therefore, we will proceed our discussion of thermodynamic properties as
a function of temperature, chemical potential and electronic density. 
Assuming that we are considering only repulsive onsite Coulomb interaction ($U>0$ ) as well as repulsive Coulomb interaction between nearest neighbor  ($V>0$). Besides $V<U$, in order to make more realistic our discussion.

\subsection{The electronic density}

We will explore the electronic density $\displaystyle
\rho=-\frac{\partial f}{\partial\mu}$, as a function of
chemical potential as well as the hopping term. In FIG.~\ref{fig:E-d
mu-t}, we plot the chemical potential $\mu/U$ versus $t/U$, for
fixed value of temperature $T/U=0.01$ and nearest neighbor coupling $V/U=0.1$. With gray scale, we mean
electronic density between $0<\rho<2$, the darkest region
corresponds to the lowest electronic density, while the brightest
region corresponds to fully filled electrons.

\begin{figure}
\includegraphics[scale=0.22]{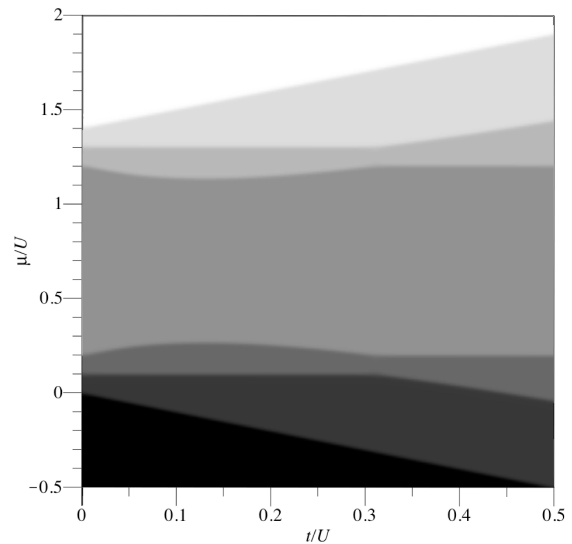}\caption{\label{fig:E-d mu-t}Electronic density per site for  $T/U=0.01$ and $V/U=0.1$, as
a function of the $t/U$ and $\mu/U$, where the black region corresponds
to empty particles, while the white region corresponds to fully filled
electrons or empty holes, by different levels of gray regions we represent
the intermediate electronic density.}
\end{figure}

\begin{figure}$
\begin{array}{c}
\includegraphics[scale=0.35]{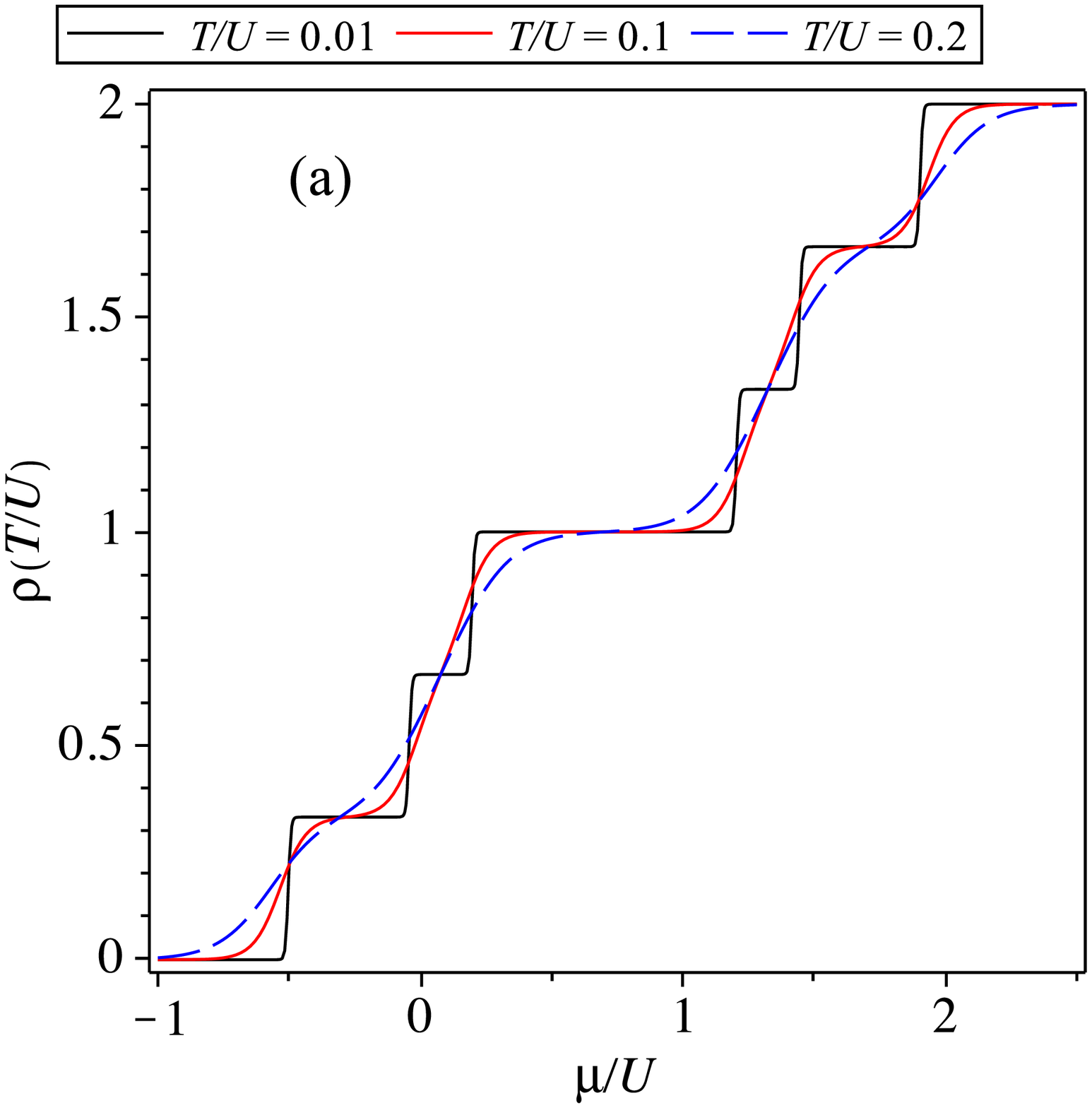}
\\
\\\includegraphics[scale=0.35]{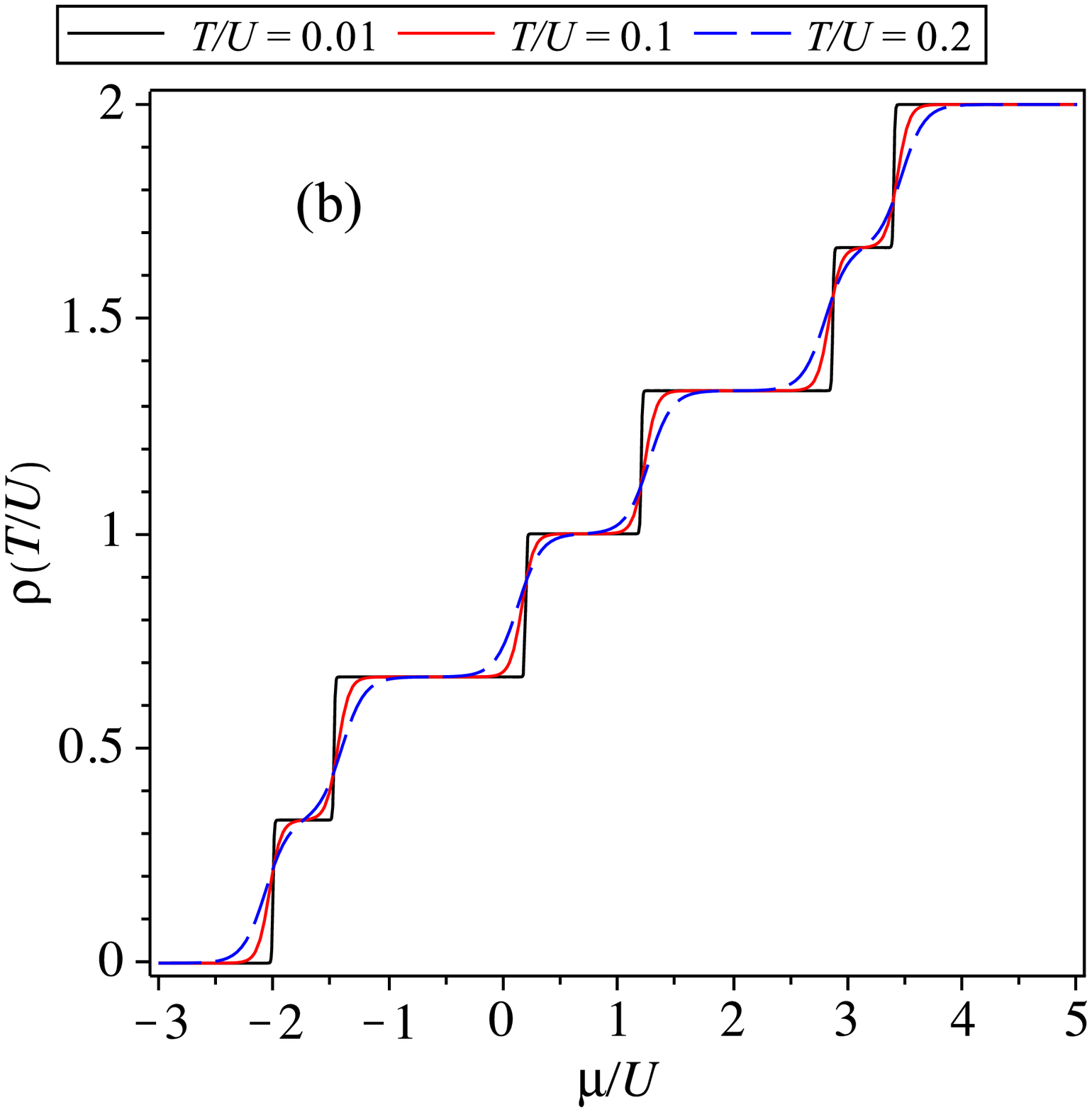}

\end{array}$\caption{\label{fig:E-d mu fixed t}(Color online) Electronic density as a function of chemical
potential, for fixed value of $V/U=0.1$. (a) For $t/U=0.5$ and in (b) for $t/U=2.0$}
\end{figure}

In FIG.~\ref{fig:E-d mu fixed t}(a), we plot the electronic density
as a function of chemical potential at low temperature for fixed
value of hopping term $t/U=0.5$ and $V/U=0.1$, showing six plateaus at $\rho=0$, $1/3$,
$2/3$, $1$, $4/3$, $5/3$ and $2$, this phenomenon vanishes as soon as
the temperature increases. Whereas in FIG.~\ref{fig:E-d mu fixed
t}(b) we plot the same quantity but for large hopping term $t/U=2$ and $V/U=0.1$.
In this case the plateaus corresponding to electronic densities
$\rho=1/3$, $2/3$, $4/3$ and $5/3$, which turn away from the half
filled band $\rho=1$. Moreover, the plateaus for densities $2/3$ and
$4/3$ become larger, as we can show in FIG.~\ref{fig:E-d mu fixed
t}, which is also in agreement with FIG.~\ref{fig:E-d mu-t}.

\subsection{The internal energy}

\begin{figure}$
\begin{array}{c}
\includegraphics[scale=0.35]{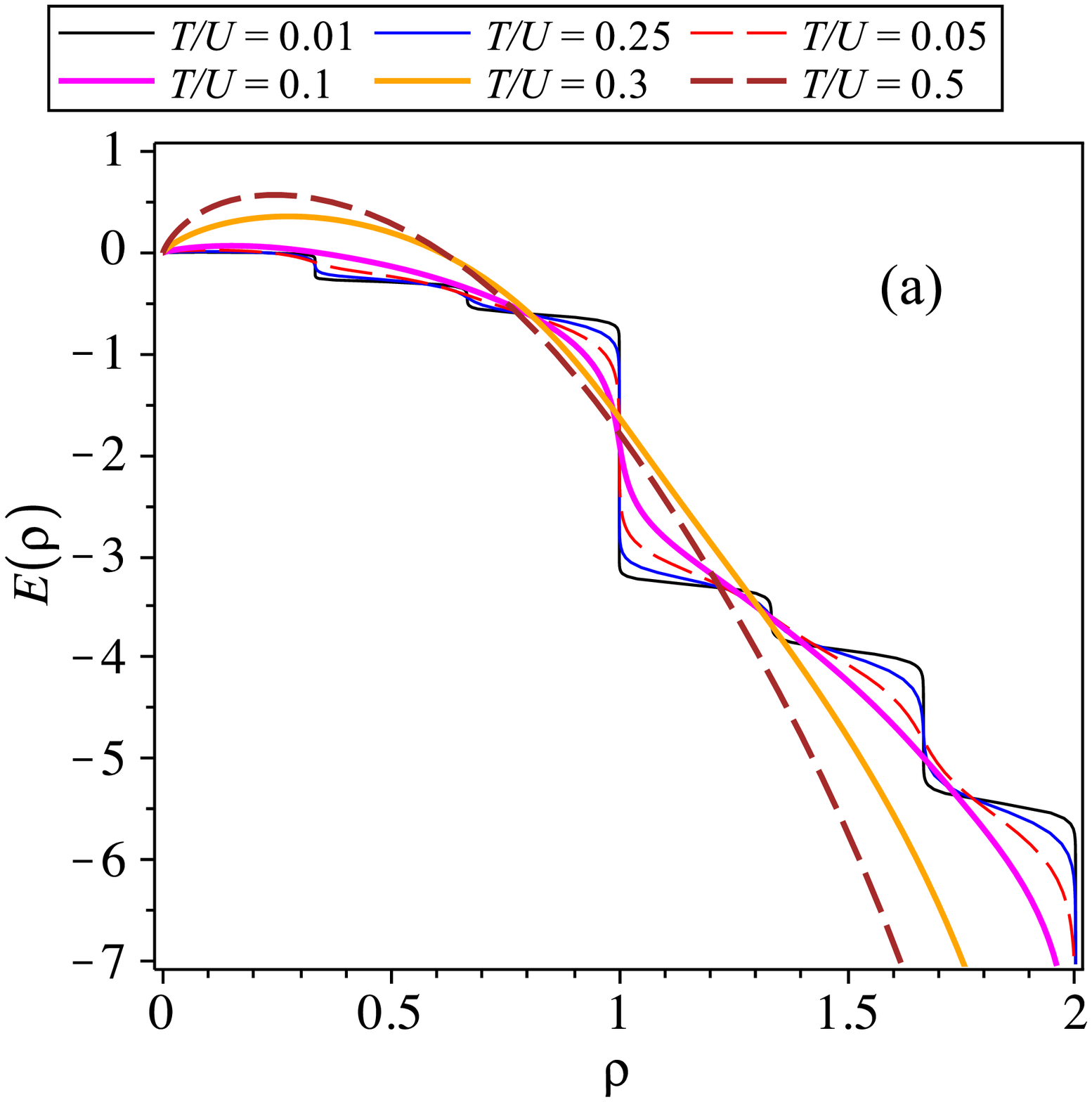}
\\
\includegraphics[scale=0.35]{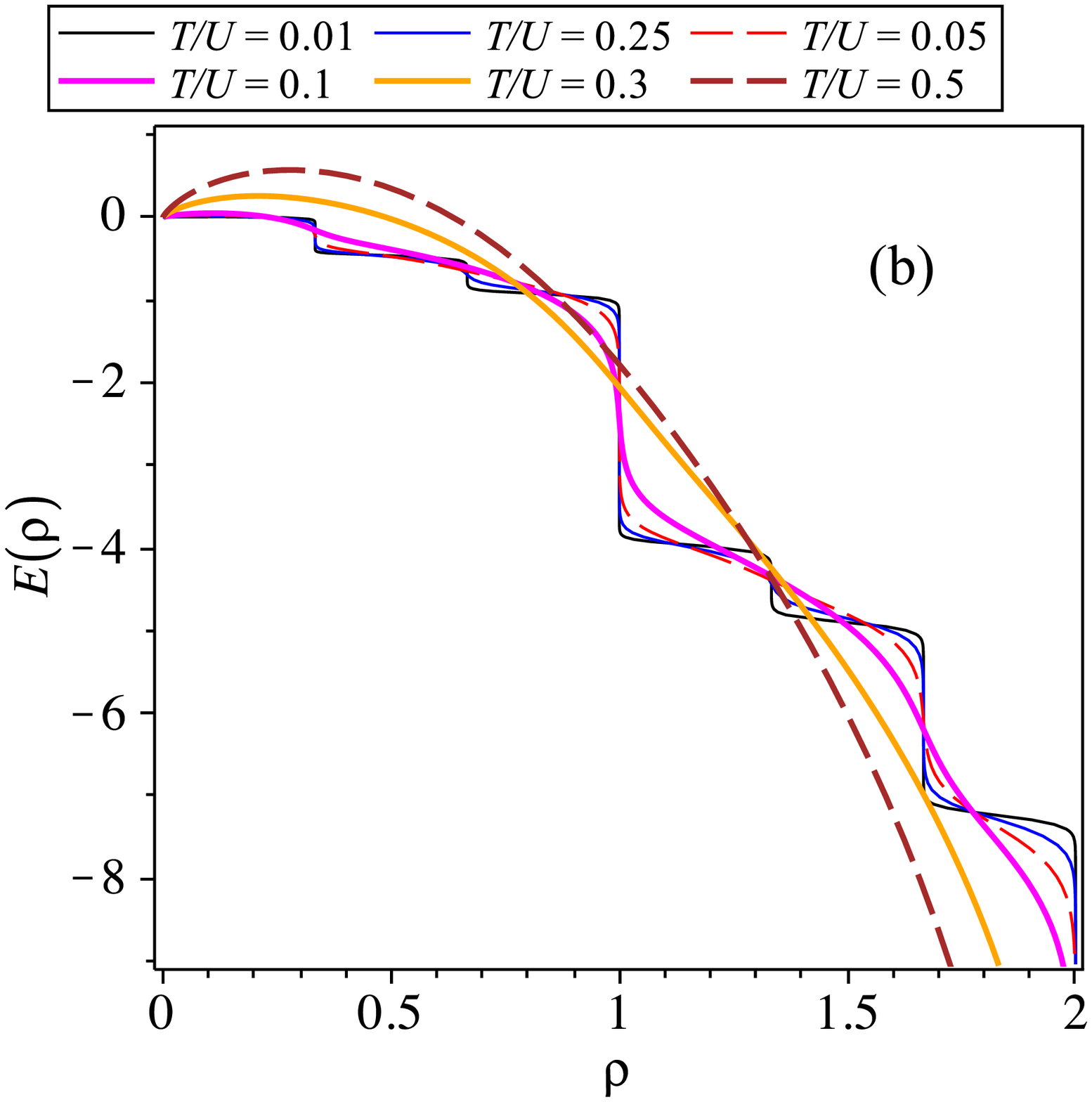}

\end{array}$\caption{\label{fig:U-energy}(Color online) Internal energy $\mathscr{E}(\rho)$ as a function
of electronic density $\rho$ for given value of $V/U=0.1$. (a) For $t/U=0.2$ and in (b) for $t/U=0.5$ }
\end{figure}

The internal energy $\displaystyle
\mathscr{E}=-\frac{\partial\ln(\Lambda_{0})}{\partial\beta}$, for
extended Hubbard model will be discussed as a function of electronic
density. In FIG.~\ref{fig:U-energy}(a) the internal energy is
plotted  for small hopping term $t/U=0.2$  and $V/U=0.1$, this internal energy
exhibits a gap energy (step like function) at zero temperature, when the electronic
density changes, but when temperature increases the step like function becomes smoother. Whereas in FIG.~\ref{fig:U-energy}(b) for  hopping term $t/U=0.5$ and $V/U=0.1$ is plotted and the shape of curves behaves basically similar plot to that  $t/U=0.2$. In summary, FIG.~\ref{fig:U-energy} displays the electronic density dependence as well as temperature
dependence. We can see as far as the larger is the electronic
density, the lower becomes the energy, although for low density around $\rho=0.5$ a maximum of internal energy.
We  also notice for several fixed values of temperature, the plot for the
internal energy becomes smooth for sufficiently  temperature.

\subsection{Compressibility}

\begin{figure}$
\begin{array}{c}
\includegraphics[scale=0.35]{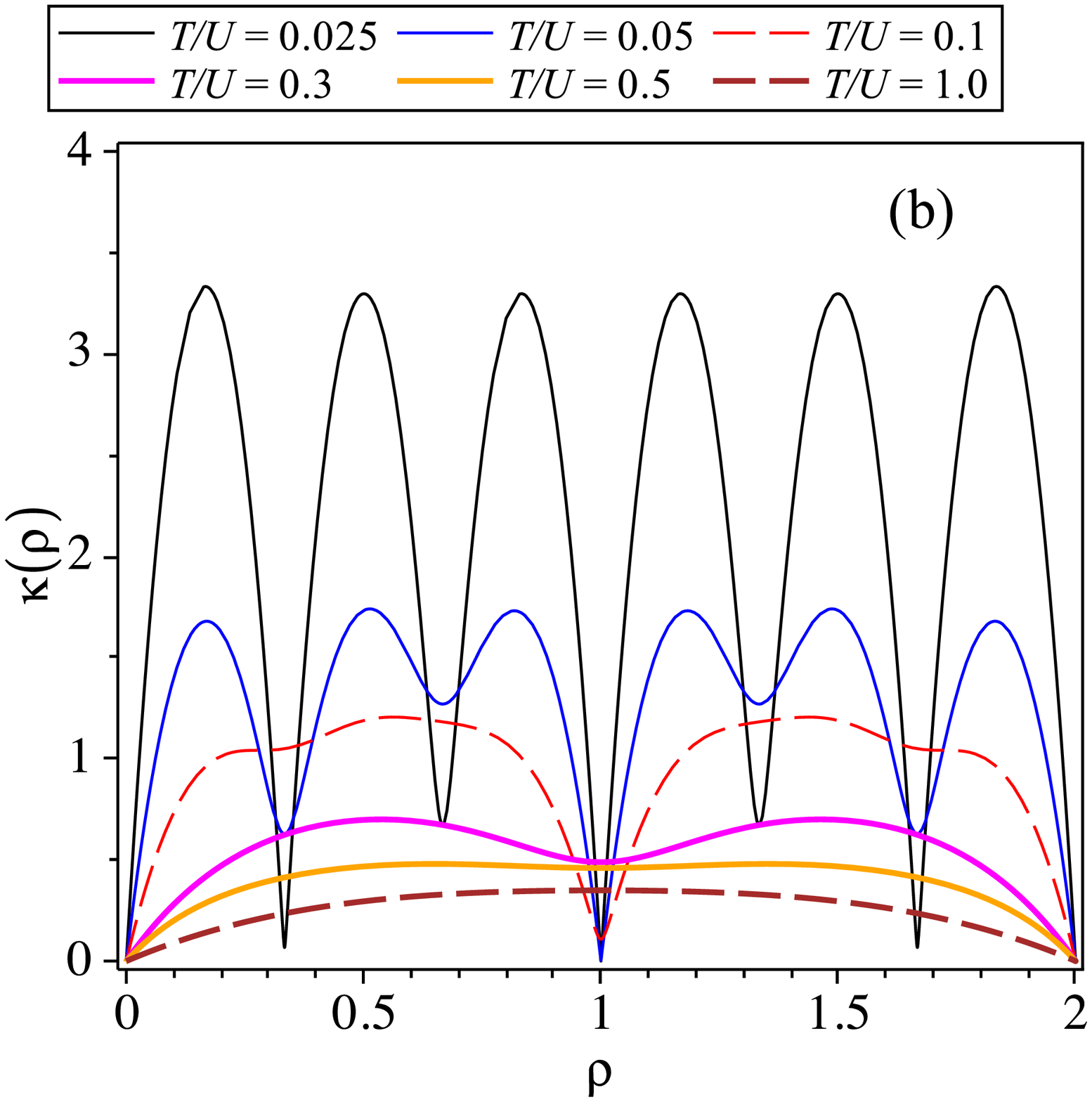}
\\
\includegraphics[scale=0.35]{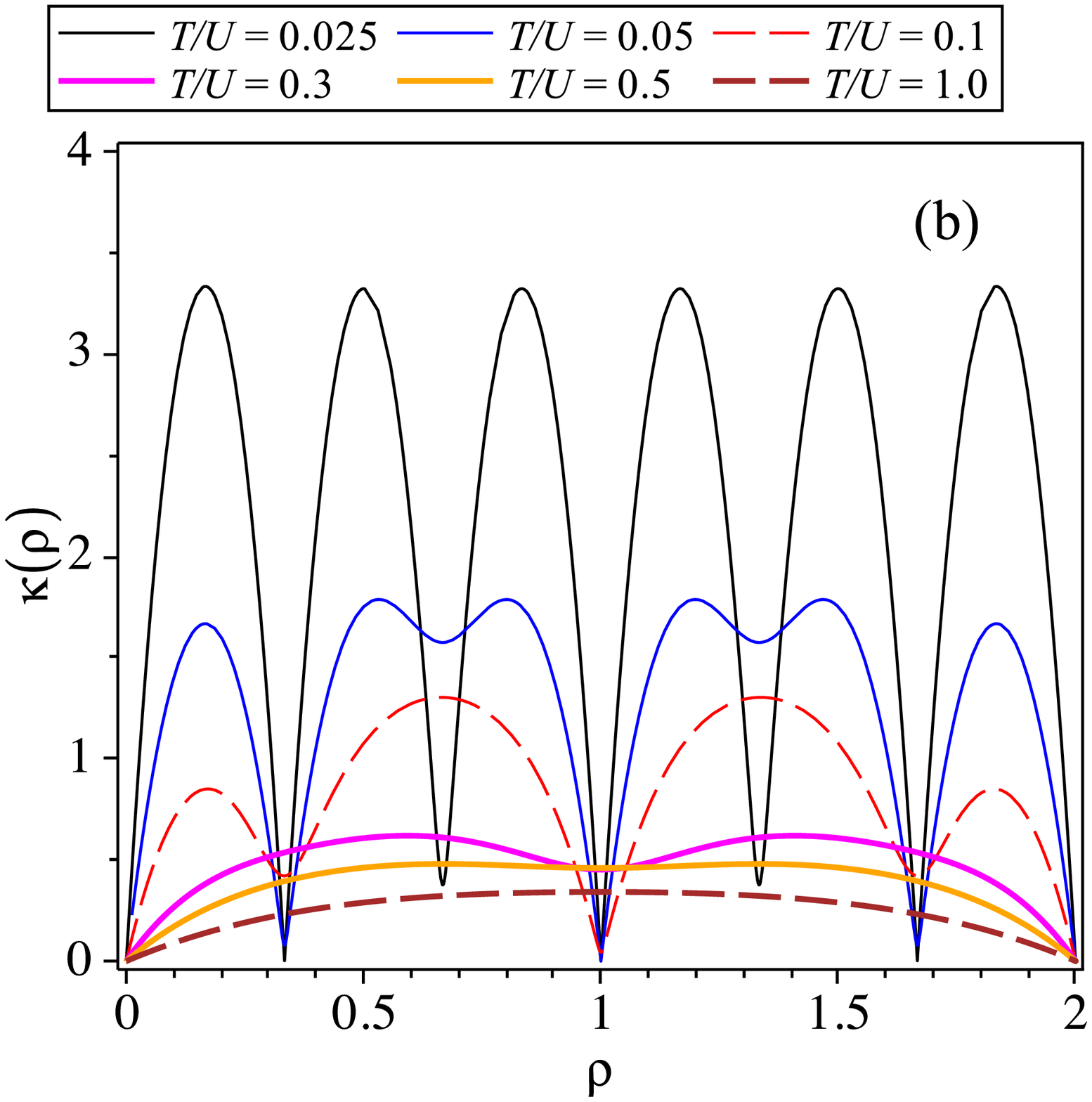}
\end{array}$\caption{\label{fig:Com-rho}(Color online) The compressibility $\kappa(\rho)$ as a function
of electronic density $\rho$, for a fixed values of $V/U=0.1$. (a) For
$t/U=0.2$, and in (b) for $t/U=0.5$. }
\end{figure}

Another interesting amount that we could discuss will be the
compressibility defined by $\kappa(\rho)\equiv\frac{\partial\rho}{\partial \mu}$, which is convenient to study electron particles, instead of that compressibility sometimes known as total compressibility\cite{baxter-book}  $
\kappa_T=-\frac{1}{\rho^{2}}\frac{\partial^{2}f}{\partial\mu^{2}}=\displaystyle \frac{\kappa}{\rho^2}$.  Therefore the compressibility  will be discussed  as a function of Hamiltonian parameters, temperature and electronic density. In FIG.~\ref{fig:Com-rho} we
display the compressibility behavior against the electronic density
on the diamond chain, for fixed value of $V/U=0.1$. In FIG.~\ref{fig:Com-rho}(a) we fixed the
hopping term at $t/U=0.2$, roughly below the phase transition occurs
at $t/U=0.3$, where we  display how the compressibility behaves at low
temperature. The compressibility becomes minimum (harder) close to the
following fractional electronic densities, $0$,  $1/3$, $2/3$, $1$, $4/3$,
$5/3$ and $2$. Whereas there are 6 local maximum (softer) between fractional
densities $1/3$, $2/3$, $1$, $4/3$, $5/3$ and $2$, for temperature
lower than $T/U\lesssim0.05$. Whereas for $0.05\lesssim T/U\lesssim 0.5$, there is just one minimum at $\rho=1$, besides the minimum at $\rho=0$ and $\rho=2$. Furthermore, for higher temperature $T/U\gtrsim0.5$ the
compressibility exhibits just one just one maximum at electronic density  $\rho=1$. In FIG.~\ref{fig:Com-rho}(b) we illustrate for $t/U=0.5$,
roughly above the phase transition. In principle, the behavior is
quite similar to FIG.~\ref{fig:Com-rho}(a), a low compressibility survives
for higher temperature $t/U=0.1$ at half filled band $\rho=1$.
Although, when temperature is sufficiently high ($0.3\lesssim T/U\lesssim 0.5$) there are only one
minimum at $\rho=1$, besides empty (fully) electrons, whereas for  $T/U\gtrsim0.5$, showing just a simple maximum at $\rho=1$.

\subsection{The entropy }
\begin{figure}$
\begin{array}{c}
a)\quad T/U=0.01\\
\includegraphics[scale=0.22]{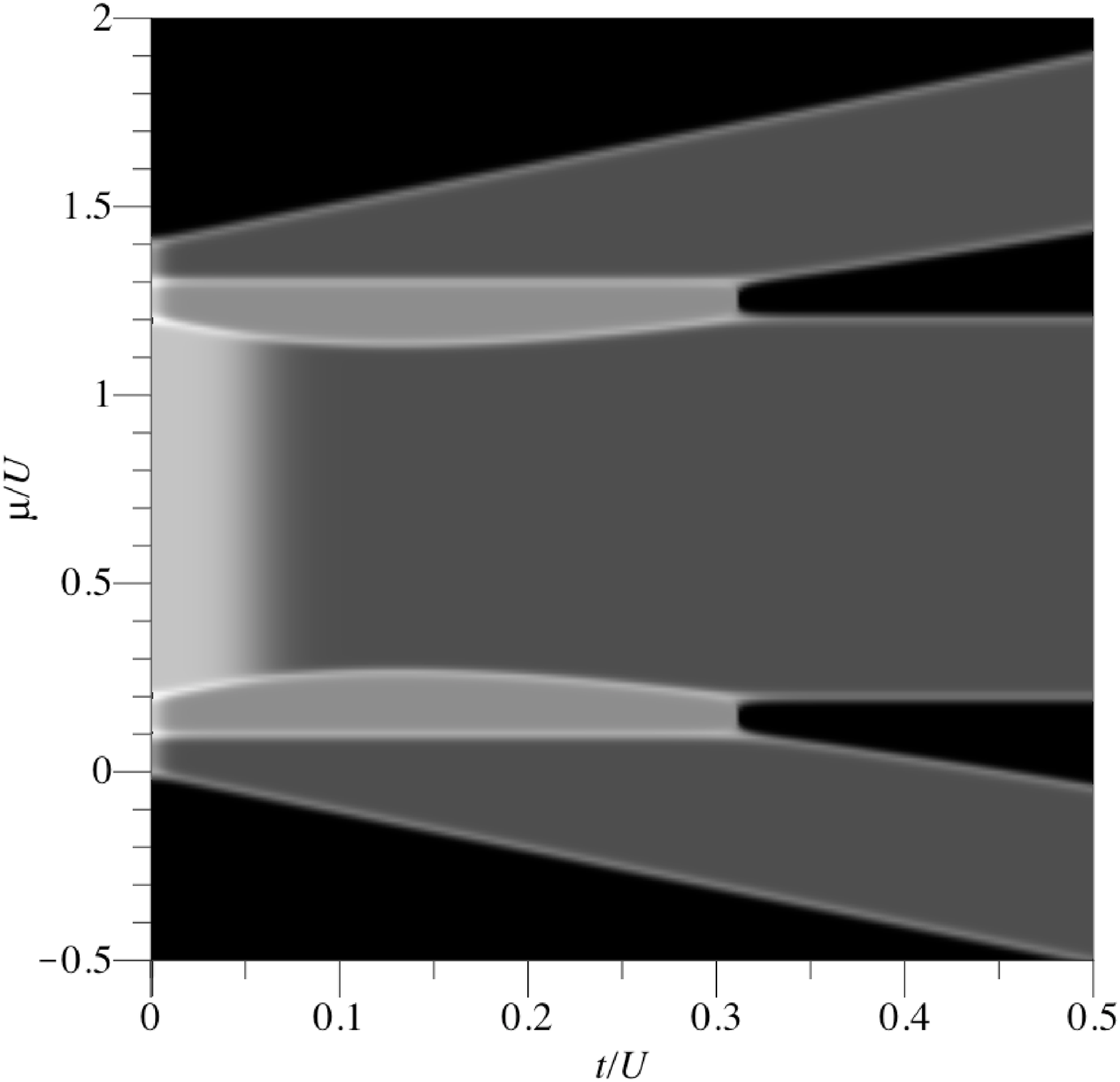}
\\b)\quad T/U=0.05
\\
\includegraphics[scale=0.22]{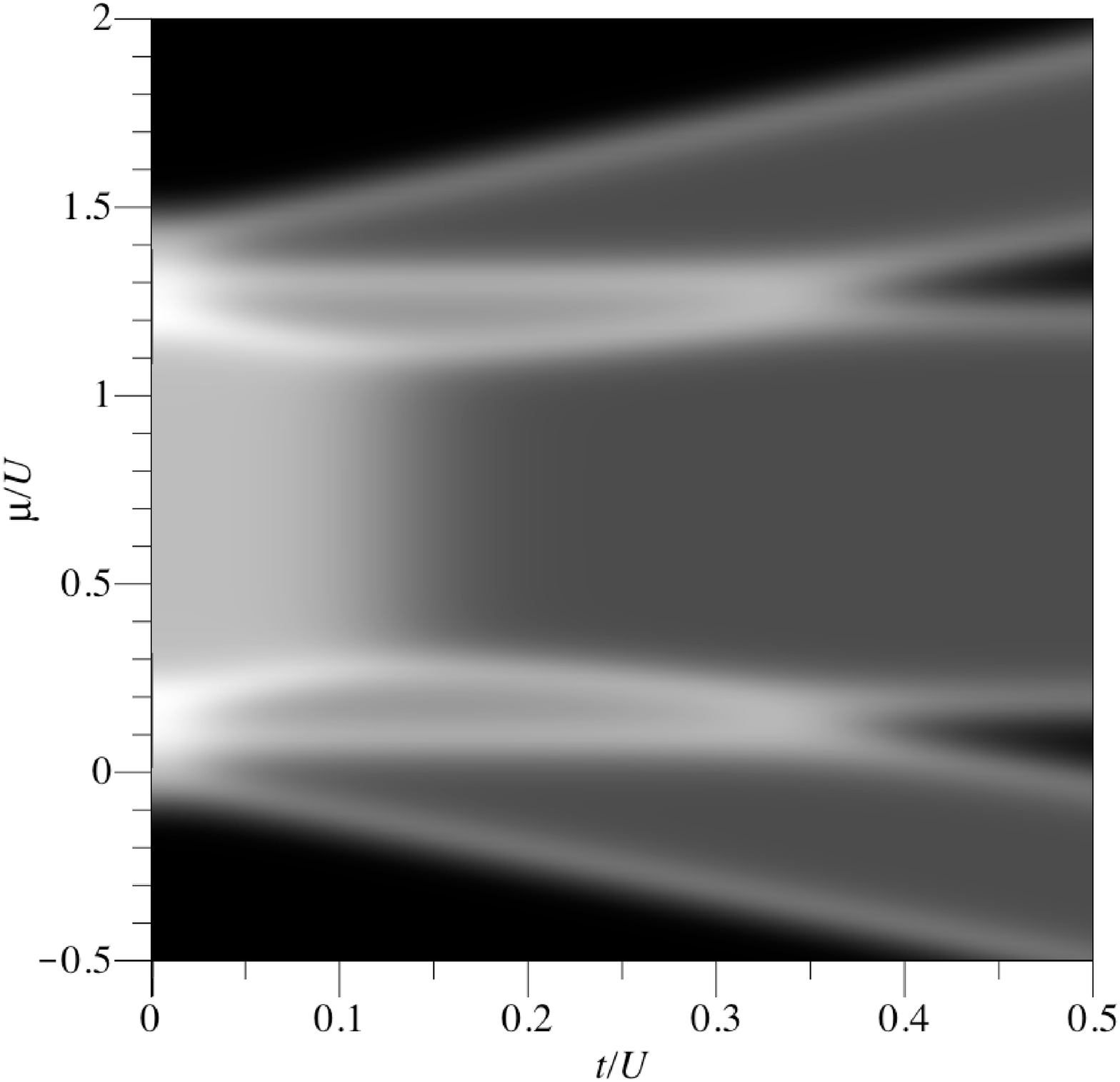}
\end{array}$
\caption{\label{fig:Res-entrop}Entropy as a function of $t/U$ and
$\mu/U$, for fixed value of $U/V=0.1$. Black region corresponds to non-frustrated entropy while
the white region correspond to higher residual entropy effect, gray
states correspond to  intermediate residual entropy effect.}
\end{figure}

Furthermore, we will study another interesting quantity the entropy $\displaystyle \mathcal{S}=-\frac{\partial f}{\partial T}$, and how the entropy behaves when the Hamiltonian parameters, temperature or even electronic  density changes.
In FIG.~\ref{fig:Res-entrop} we illustrate the magnitude of entropy, as a
function of $t/U$ and $\mu/U$, assuming a fixed value of $V/U=0.1$. By different levels of gray scale, we
represent the magnitude of entropy, the darkest region corresponds
to lowest entropy while the brightest region corresponds to the
higher entropy. At low temperature, the large value of entropy is
related to the influence of the residual entropy. As far the
temperature increases, the entropy curves become soften, as
displayed in FIG.~\ref{fig:Res-entrop}(a) for $T/U=0.01$ and in
FIG.~\ref{fig:Res-entrop}(b) for $T/U=0.05$.

\begin{figure}$
\begin{array}{c}
\includegraphics[scale=0.35]{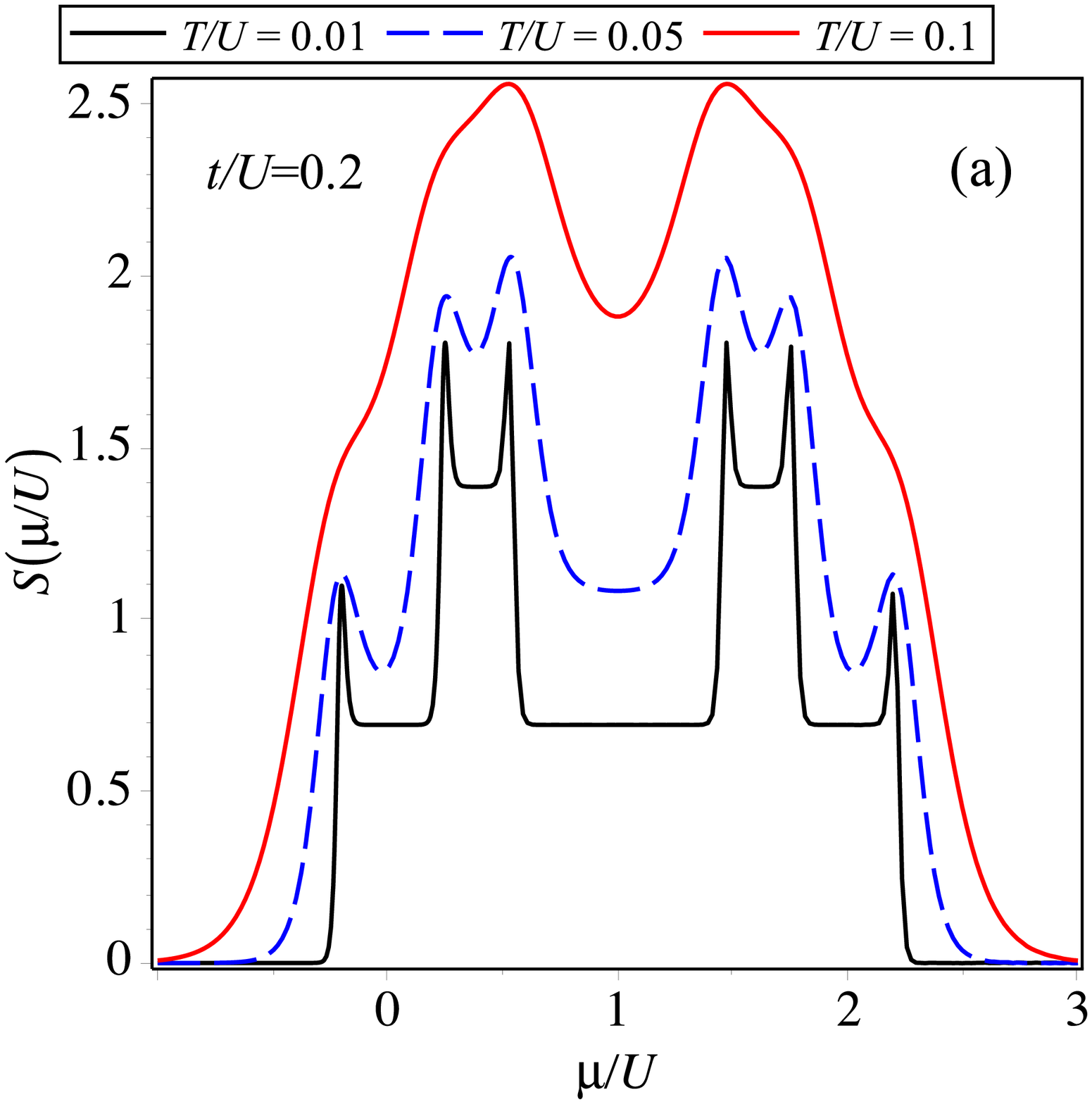}
\\
\includegraphics[scale=0.35]{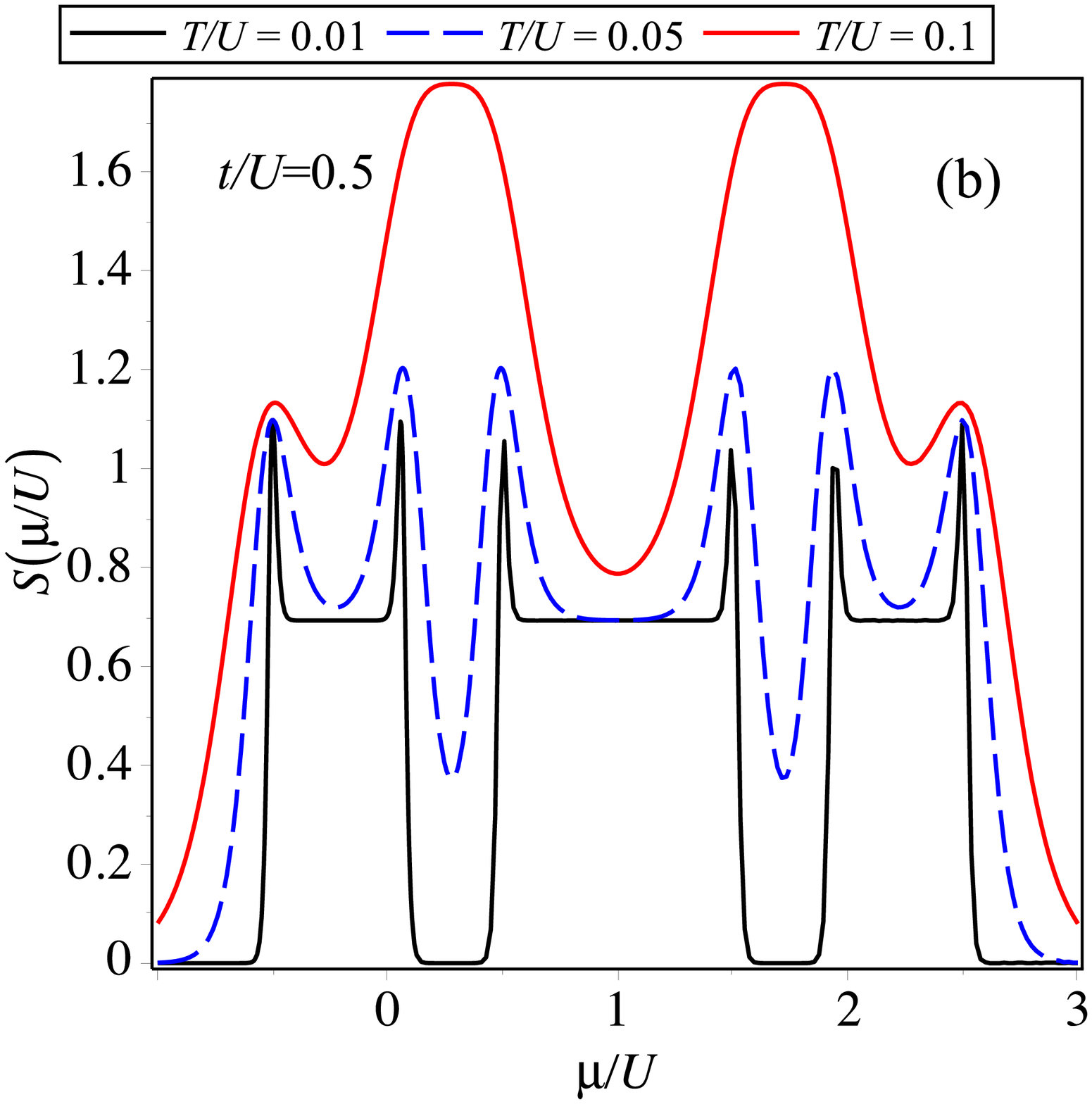}
\end{array}$\caption{\label{fig:Entropy--mu}(Color online) Entropy $\mathcal{S}$ against $\mu$ at low
temperature assuming a fixed value of $V/U=0.25$. (a) For $t/U=0.2$, and in (b) for $t/U=0.5$}

\end{figure}

Mean while in FIG.~\ref{fig:Entropy--mu}, we display the entropy
$\mathcal{S}(\mu)$ as a function of chemical potential, assuming a fixed value of $V/U=0.25$, where we are
able to illustrate the residual entropy effect, in agreement to FIG.~\ref{fig:Res-entrop}.

The entropy indicates several peaks via the chemical potential. In
FIG.~\ref{fig:Entropy--mu}(a) we plot for $t/U=0.2$ where is shown the
influence of two types of residual entropy around of
$\mathcal{S}=\ln(2)$ and $\ln(3)$. Meanwhile, in
FIG.~\ref{fig:Entropy--mu}(b) we display for $t/U=0.5.$ In this case we
only display the influence of the residual entropy of
$\mathcal{S}=\ln(2)$, certainly with the temperature increases we
have the increase of entropy as well, however for large $|\mu|$ the
entropy is smaller than for small $\mu\approx1$.

\begin{figure}$
\begin{array}{c}
\includegraphics[scale=0.35]{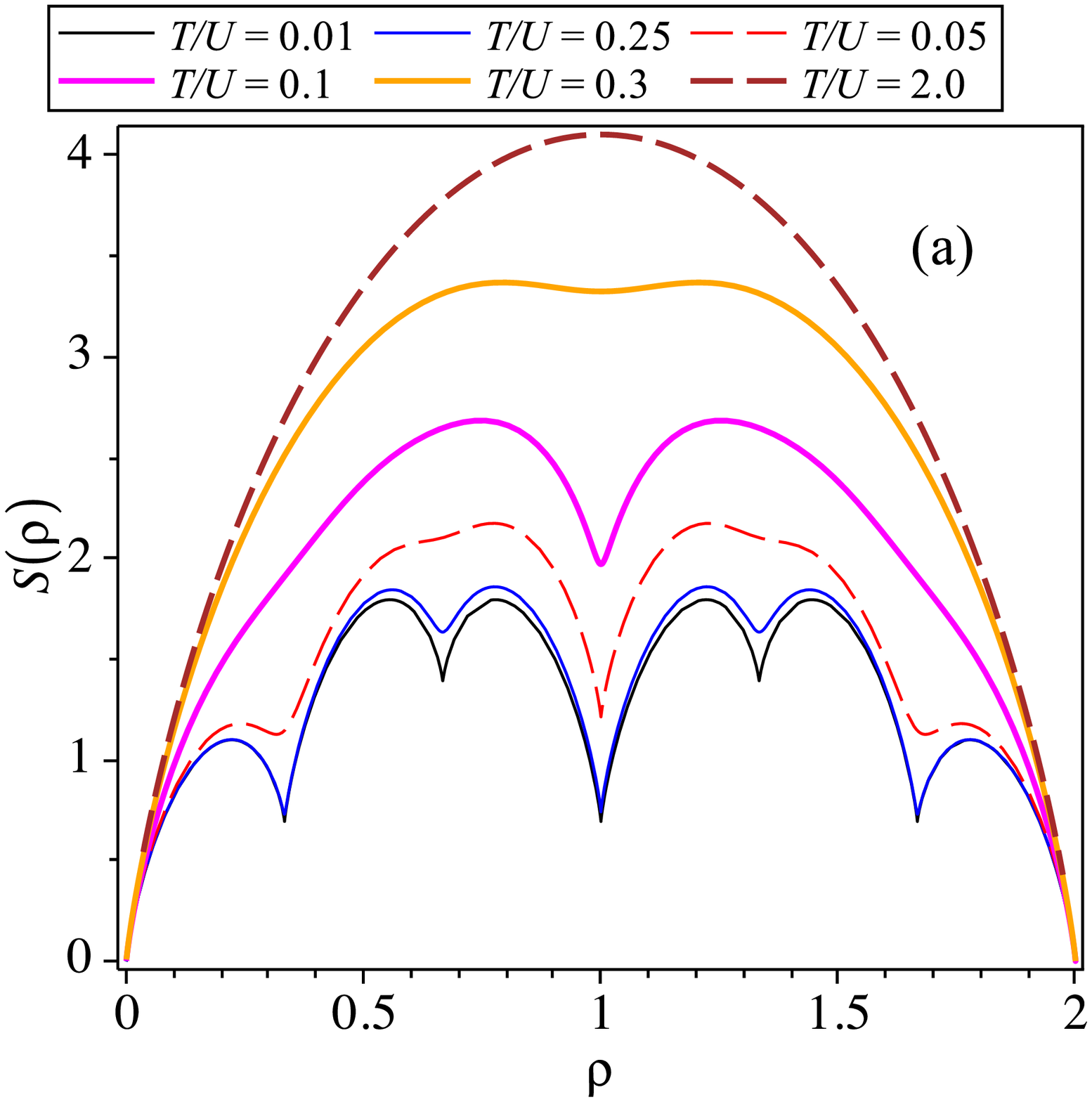}
\\
\includegraphics[scale=0.35]{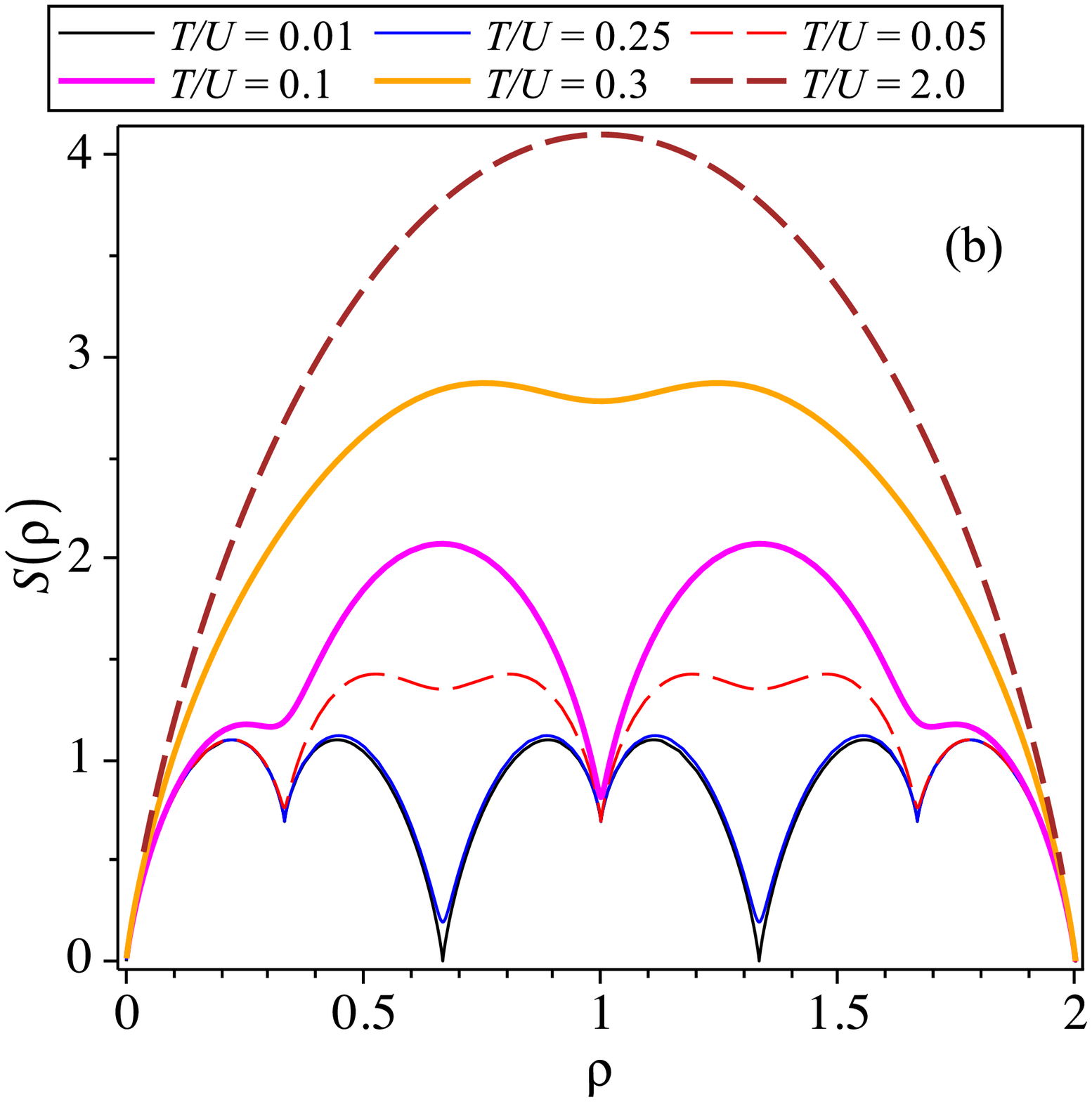}

\end{array}$\caption{\label{fig:Entropy-rho}(Color online) Entropy against electronic density, assuming fixed Coulomb coupling $V/U=0.1$. (a) For
$t/U=0.2$, and in (b) for $t/U=0.5$. }
\end{figure}

Whereas in FIG.~\ref{fig:Entropy-rho}, we display the entropy as a function
of electronic density assuming a given parameter $V/U=0.1$, for fixed value of parameter $V/U=0.1$. For high temperature $T\geqslant2,$ the
entropy reaches its highest point at $\rho=1$. We can note the
residual entropy is strongly related to the electronic density. We
also indicate, there is no residual entropy for a fully electronic
density, therefore the entropy becomes lower even for high
temperature.

For electronic density $\rho=1/3$ and $2/3$ the residual entropy
is $\mathcal{S}=\ln(2)$, while for $\rho=4/3$ and $5/3$ larger
residual entropy is displayed $\mathcal{S}=\ln(4)$.

\subsection{The specific heat}
\begin{figure}$
\begin{array}{c}
\includegraphics[scale=0.35]{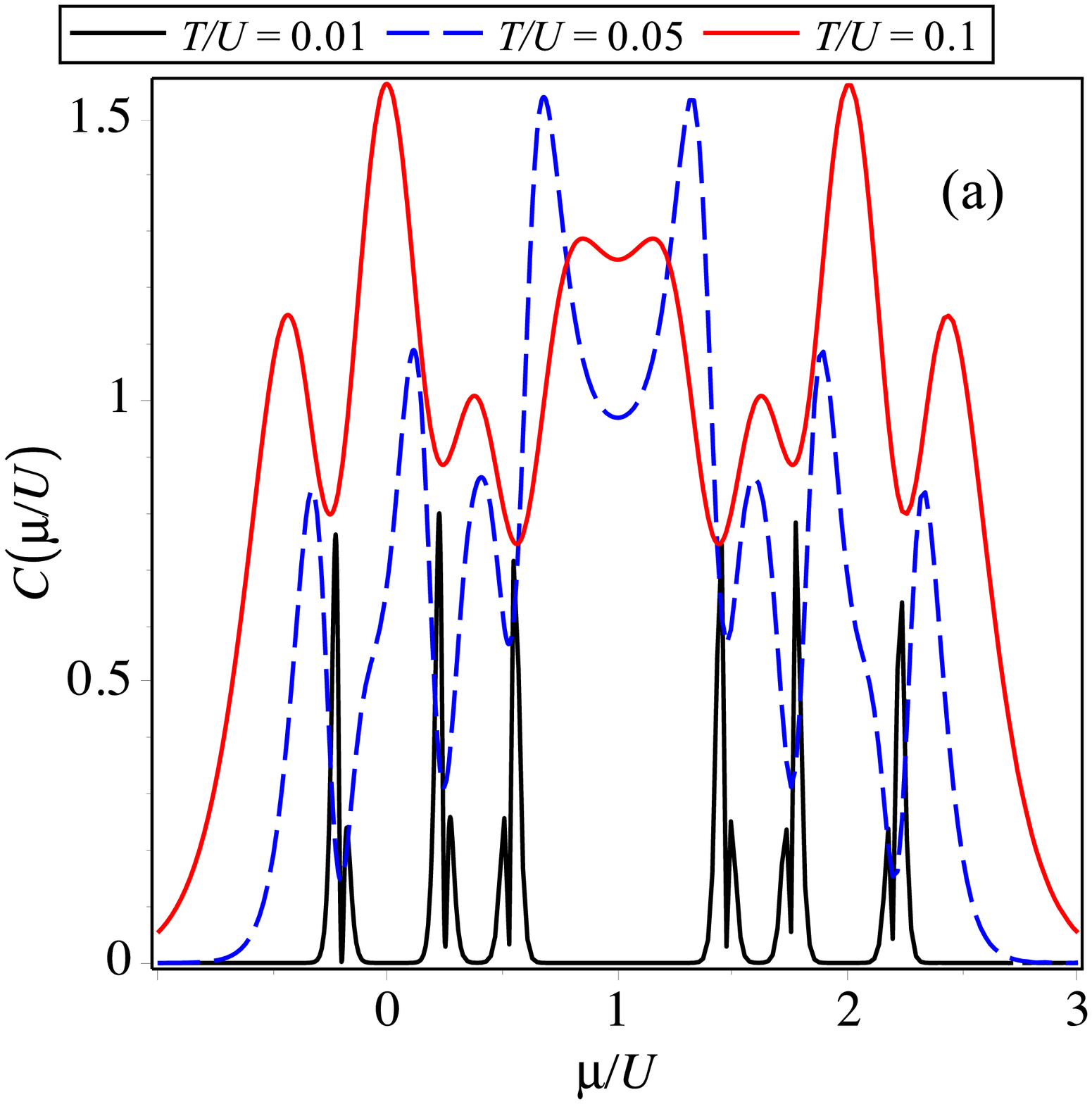}
\\
\includegraphics[scale=0.35]{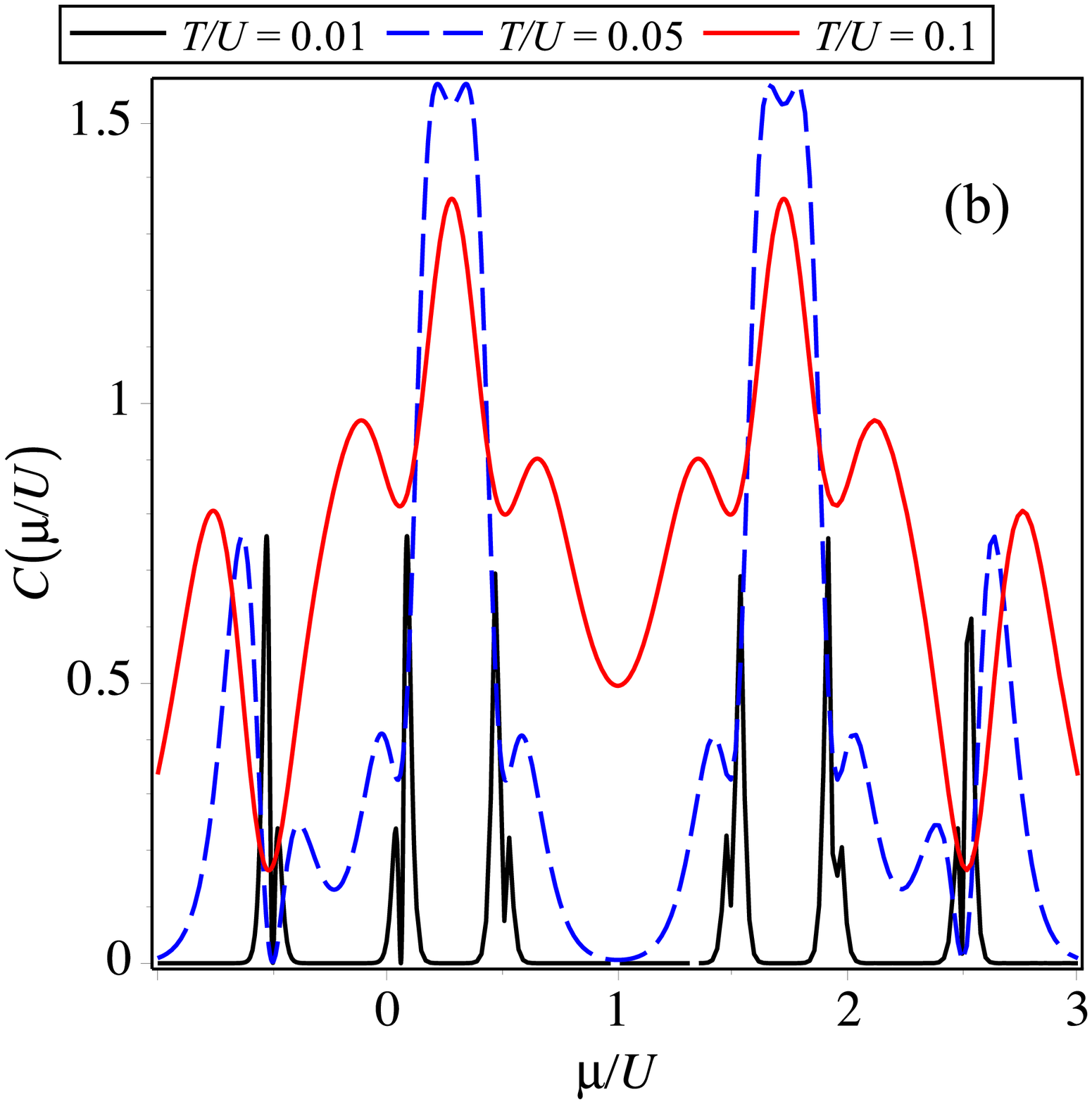}
\end{array}$
\caption{\label{fig:spc-ht vs mu}(Coloronline) The specific heat against chemical potential
for low temperature, for a fixed value of $V/U=0.25$. (a) For $t/U=0.2$, and in (b) for $t/U=0.5$}

\end{figure}

Finally we will discuss the specific heat $\displaystyle C=-T\frac{\partial^{2}f}{\partial T^{2}}$ behavior  for extended Hubbard model in quasi-atomic limit. Thus, let us start displaying the specific heat as a
function of chemical potential, which is illustrated   in FIG.~\ref{fig:spc-ht vs mu}  considering a fixed value of $V/U=0.25$.  In the low temperature limit, we
can see the effect of the phase transition at zero temperature with
very sharp peaks when the temperature decreases. Particularly, in
FIG.~\ref{fig:spc-ht vs mu}(a) we display the specific heat as a
function of $\mu/U$ for $t/U=0.2$ while in (b)  the specific heat as a
function of $\mu/U$ is illustrated for $t/U=0.5$.

\begin{figure}$
\begin{array}{c}
\includegraphics[scale=0.35]{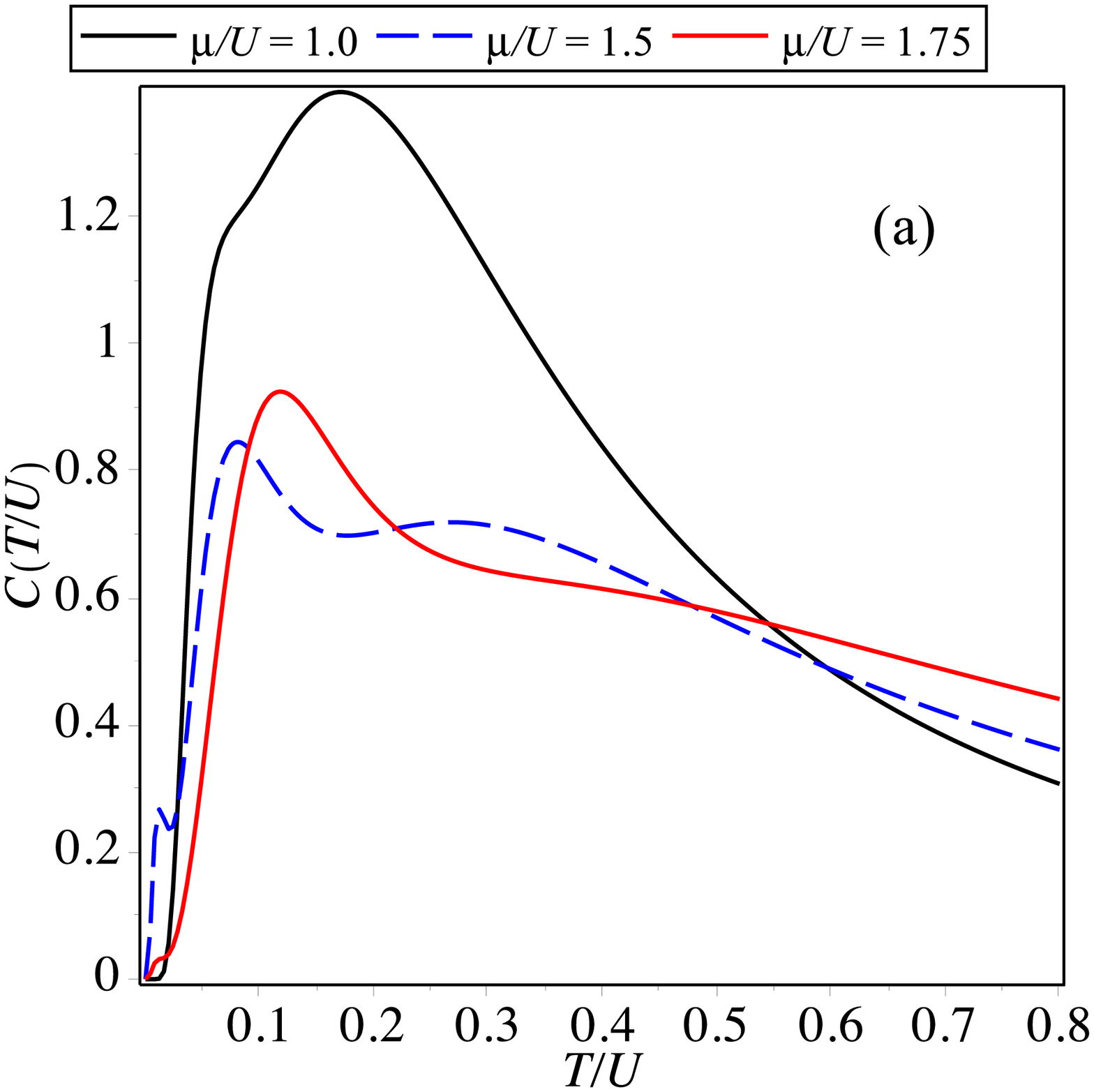}
\\
 \includegraphics[scale=0.35]{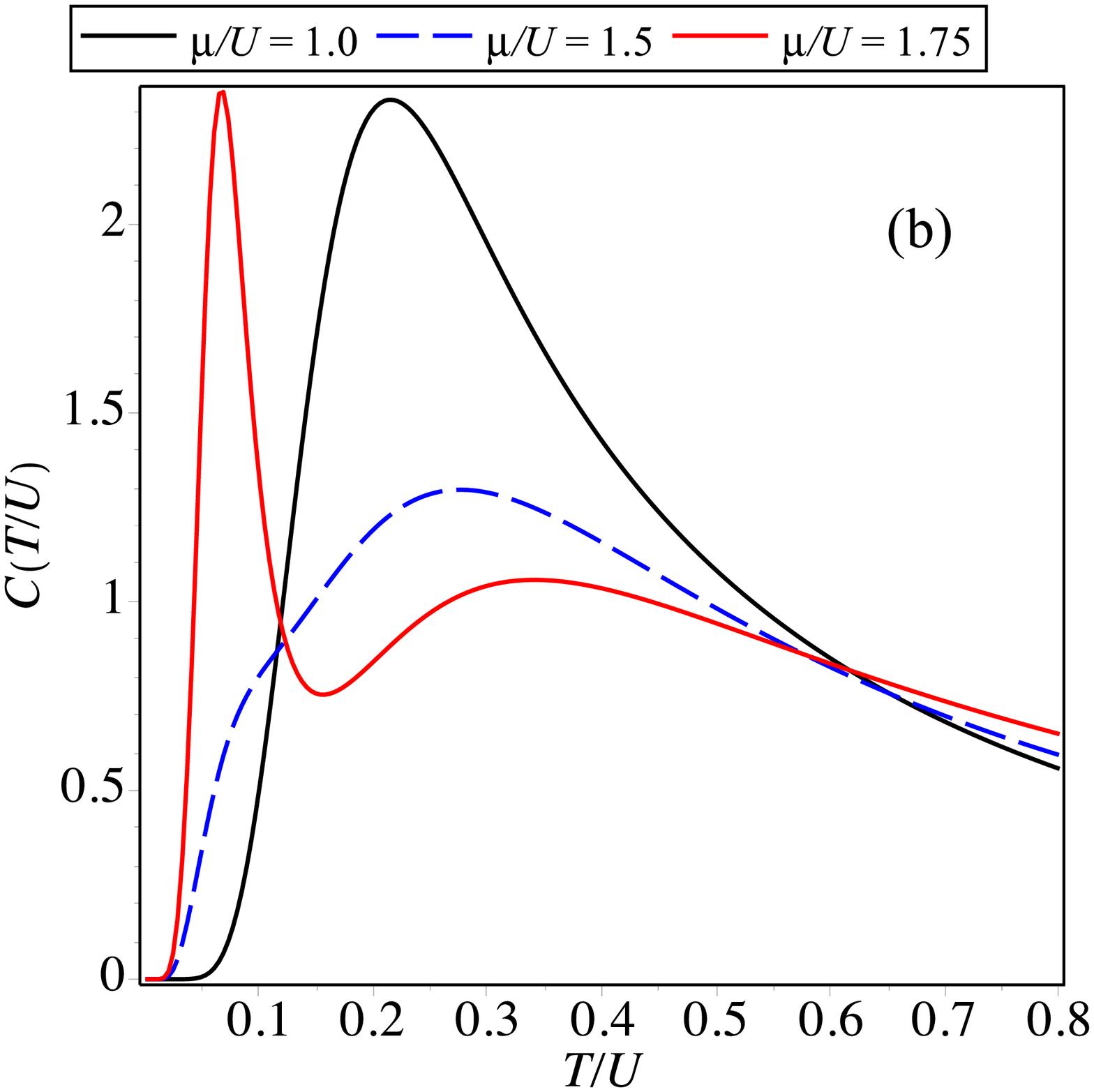}

\end{array}$
\caption{\label{fig:spc-ht vs T}(Coloronline) The specific heat versus temperature,  for a given value of $V/U=0.25$ and 
different values of chemical potential. In
(a) is considered for $t/U=0.2$, while  in (b) for $t/U=0.5$. }
\end{figure}

Lastly in FIG.~\ref{fig:spc-ht vs T} we display the specific heat as a
function of temperature assuming the nearest neighbor Coulomb interaction $V/U=0.25$. In
FIG.~\ref{fig:spc-ht vs T}(a) we consider for $t/U=0.2$, while in
FIG.~\ref{fig:spc-ht vs T}(b) we assume for $t/U=0.5$. From where we can see  very sharp
peaks as the temperature decreases, this anomalous peaks appears due to low lying energy around first order phase transition at zero temperature.

\section{\label{sec:conclusions}Conclusions}

The proposed Hubbard model on diamond chain, was discussed at zero
temperature as well as at finite temperature. The phase diagram at
zero temperature displays 4 frustrated states and 5 non-frustrated
states antiferromagnetically ordered. Concerning to finite
temperature properties, this model can be solved exactly through
decoration transformation \cite{Syozi,Fisher,phys-A-09,JPA-11}.
Transfer matrix technique \cite{baxter-book} presents mapping the
proposed model onto an exactly Hubbard model in atomic limit with
three and four body couplings. Therefore, detailed thermodynamic
properties were discussed, such as density as a function of chemical
potential, illustrating 6 plateaus. Internal energy was also
considered as function of electronic density far away from the half
filled band, showing for several fixed temperatures. Besides we
studied the compressibility. Once again we illustrate this amount as
a function of electronic density, and so   we can conclude that the
diamond chain is more compressible at low temperature, when the
electronic density is respectively between 1/3, 2/3, 1, 4/3, 5/3. As
soon as the temperature increases,  the compressibility decreases,
turning into a simple decreasing curve as a function of density. We
also considered   the entropy as another interesting amount. This
amount was studied as a function of chemical potential as well as
electronic density, where we clearly  observe the residual entropy
contribution, owing to geometric frustration. Finally we  discuss
the specific heat, as a function of chemical potential and the
temperature.  In the nearest future, we plan to study the Coulomb
interaction between the nodal (the monomer-monomer) sites and a
different type of interaction
 between the nodal-interstitial (dimer-monomer) sites, both positive or negative
ones, at the quantum level, which explains the wide range of experiments on the diamond chain.
\section*{Acknowledgment}
O. R. and de Souza thanks CNPq and Fapemig for partial financial support.
This work has been supported by the French-Armenian grant No. CNRS IE-017 (N.A.) and by the Brazilian FAPEMIG grant No. CEX â BPV â 00028-11 (N. A.).





\begin{thebibliography}{99}
\bibitem{elliot}B. Nachtergaele, J. P. Solovej, J. Yngvason, \textit{Condensed
Matter Physics and Exactly Soluble Models}, Selecta of E. H.~Lieb, (Springer, Berlin-Heidelberg, 2004).

\bibitem{hagemann} I. S. Hagemann et al., Phys. Rev. Lett.\textbf{
86,}  894 (2001).
\bibitem{lierop}J. van Lierop, D. H. Ryan, Phys. Rev. Lett. \textbf{86,}
 4390 (2001).
\bibitem{eggert}S. Eggert, Phys. Rev. B \textbf{54,} R9612 (1996).
\bibitem{okamoto} Okamoto and Nomura, Phys. Lett. A \textbf{169}, 433 (1992).
\bibitem{laflorencie}N. Laflorencie , D. Poilblanc, Phys. Rev.
Lett. \textbf{90,}  157202 (2003).

\bibitem{daul}S. Daul, R. M. Noack, Phys. Rev. B\textbf{ 61,}
1646 (2000).

\bibitem{Hida}K. Hida, J. Phys. Soc. Jpn. \textbf{63,} 2359 (1994).

\bibitem{Ohanyan}V. R. Ohanyan, N. S. Ananikian, Phys. Lett. A\textbf{ 307,}  76 (2003).

\bibitem{Hovhan}V. V. Hovhannisyan, N.S. Ananikian, Phys. Lett. A\textbf{ 372,} 3363 (2008).

\bibitem{Artuso}N. Ananikian, L. Ananikyan, R. Artuso, H. Lazaryan, Phys. Lett. A  \textbf{ 374, }4084 (2010).

\bibitem{Koch}A. N. Kocharian, G. W. Fernando, K. Palandage, J. W. Davenport, Phys. Rev. B \textbf{ 74,}
024511 (2006).

\bibitem{Koch2}A. N. Kocharian, G. W. Fernando, K. Palandage, T. Wang, J.
W. Davenport,
 Phys. Lett. A \textbf{ 364,} 57 (2007).

\bibitem{Derzhko09}O. Derzhko, A. Honecker, J. Richter, Phys.
Rev. B \textbf{79,}  054403 (2009).
\bibitem{Derzhko10}O. Derzhko, J. Richter, A. Honecker, M. Maksymenko,
 R. Moessner Phys. Rev. B \textbf{81,} 014421 (2010).

\bibitem{montenegro06}R. R. Montenegro-Filho,  M.D. Coutinho-Filho,
Phys. Rev. B \textbf{74,}  125117 (2006).

\bibitem{Mancini}F. Mancini, Eur. Phys. J. B \textbf{47,} 527 (2005).
\bibitem{mancini2008}
 F. Mancini,  F. P. Mancini
Phys. Rev. E \textbf{77,} 061120 (2008).
\bibitem{Vidal98-00}J. Vidal, R. Mosseri,  B. Doucot, Phys. Rev.
Lett. \textbf{81,} 5888 (1998); J. Vidal et al., \textbf{85,}  3906 (2000); B. Doucot and J. Vidal, Phys. Rev. Lett. \textbf{88}, 227005 (2002).

\bibitem{rossler} J. R\"ossler, D. Mainemer, Cond. Matter Phys.
\textbf{13,}  13704 (2010).


\bibitem{WZ Wang}W. Z. Wang, Phys. Rev. B \textbf{72,}  125116 (2005).

\bibitem{gulacsi}Z. Gul\'acsi, A. Kampf,  D. Vollhardt, Phys. Rev.
Lett. \textbf{99,} 026404 (2007).

\bibitem{gulacsi PTP}Z. Gul\'acsi, A. Kampf, and D. Vollhardt, Prog.
Theor. Phys. Supp. \textbf{176,} 1 (2008).
\bibitem{spinless}O. Rojas,  S.M. de Souza, Phys. Lett. A \textbf{375,}
 1295 (2011).
\bibitem{Lopes} A. A. Lopes, R.G. Dias, Phys. Rev. B \textbf{84,}
  085124 (2011).
\bibitem{honecker}A. Honecker, S. Hu, R. Peters  J. Ritcher, J.
Phys.: Condens. Matter \textbf{23,} 164211 (2011).


\bibitem{Pereira08}M. S. S. Pereira, F. A. B. F. de Moura, M.
L. Lyra, Phys. Rev. B \textbf{77,}  024402 (2008).

\bibitem{pereira09}M. S. S. Pereira, F. A. B. F. de Moura, M.
L. Lyra, Phys. Rev. B \textbf{79,} 054427 (2009).

\bibitem{lisnii}B. M. Lisnii, Low Temp. Phys. \textbf{37,} 296 (2011).

\bibitem{strecka prb}J. Stre\v{c}ka, A. Tanaka, L. \v{C}anova,
T. Verkholyak, Phys. Rev. B \textbf{80,}  174410 (2009).

\bibitem{strecka conf}J. Stre\v{c}ka, A. Tanaka, M. Jas\v{c}ur Journal
of Physics Conference Series \textbf{200,}  022059 (2010).

\bibitem{canova06}L. Canova, J. Strecka, M. Jascur, J. Phys.: Condens.
Matter \textbf{18,}  4967 (2006).

\bibitem{vadim}O. Rojas, S.M. de Souza, V. Ohanyan, M. Khurshudyan,
Phys. Rev. B \textbf{83} , 094430 (2011).

\bibitem{valverde}J. S. Valverde, O. Rojas, S. M. de Souza, J.
Phys. Condens. Matter\textbf{ 20,}  345208 (2008).
\bibitem{lisnii-11}B. M. Lisnii, Ukrainian Journal of Physics \textbf{56,}
1237 (2011).

\bibitem{ananikian}N. S. Ananikian, L. N. Ananikyan, L. A. Chakhmakhchyan and O. Rojas,
arXiv:1110.6406.
\bibitem{rule} K. C. Rule et al., Phys. Rev. Lett. \textbf{100}, 117202 (2008).

\bibitem{kikuchi}H. Kikuchi et al.,  Phys. Rev. Lett. \textbf{94}, 227201 (2005); H.~Kikuchi et al., Prog. Theor. Phys. Suppl. \textbf{159}, 1 (2005).
\bibitem{honecker-lauchli} A. Honecker and A. Lauchli, Phys. Rev. B \textbf{63}, 174407 (2001);  H. Jeschke et al., Phys. Rev. Lett. \textbf{106}, 217201 (2011); J. Kang et al., J. Phys. Condens. Matter \textbf{21}, 392201(2009); K. Takano, K. Kubo and H. Sakamoto, J.~Phys.: Condens. Matter \textbf{8}, 6405 (1996); K Takano, K.~Kubo and H. Sakamoto, J. Phys.: Condens. Matter \textbf{15}, 5979 (2003).
\bibitem{chakh}L. Chakhmakhchyan, N. Ananikian, L. Ananikyan and C. Burdik, J. Phys: Conf. Series \textbf{343}, 012022 (2012).
\bibitem{sh-deng}Sh.-S. Deng, Sh.-J. Gu, and H.-Q. Lin, Phys. Rev. B \textbf{74}, 045103 (2006).
\bibitem{Bo-Gu} B. Gu and G. Su, Phys.  Rev. B \textbf{75}, 174437 (2007); Kang~et al., J. Phys. Condens. Matter \textbf{21}, 392201(2009); O. Derzhko and J. Richter, Eur. Phys. J. B \textbf{52}, (2006)~23;  B. Gu and G. Su, Phys. Rev. Lett. \textbf{97}, 089701 (2006); H.~Kikuchi et al., Phys. Rev. Lett. \textbf{97}, 089702 (2006).
\bibitem{Syozi}I. Syozi, Prog. Theor. Phys.\textbf{ 6,} 341 (1951).

\bibitem{Fisher} M. Fisher, Phys. Rev. \textbf{113,} 969 (1959).

\bibitem{phys-A-09}O. Rojas, J. S. Valverde,  S. M. de Souza, Physica
A \textbf{388,}  1419 (2009).

\bibitem{strecka pla}J. Stre\v{c}ka, Phys. Lett. A\textbf{ 374,} 3718 (2010).


\bibitem{JPA-11}O. Rojas, S. M. de Souza, J. Phys. A: Math. Theor.
\textbf{44,}  245001 (2011).




\bibitem{m rojas}M. Rojas, O. Rojas,  S.M. de Souza  arXiv:1105.5130.



\bibitem{baxter-book}R. J. Baxter, \textit{Exactly Solved Models in
Statistical Mechanics}, (Academic Press, New York, 1982).
\bibitem{beni-pincus}G. Beni, P. Pincus Phys. Rev. B \textbf{9,}  2963 (1974).

\end{thebibliography}







\end{document}